\documentclass[12pt,titlepage]{article}

\font\grande=cmr9.5 scaled \magstep4
\font\medio=cmr9.5 scaled \magstep2
\outer\def\beginsection#1\par{\medbreak\bigskip
      \message{#1}\leftline{\bf#1}\nobreak\medskip
\vskip-\parskip
      \noindent}
\usepackage{graphicx} 
\begin{document}
\bibliographystyle {unsrt}

\titlepage

\begin{flushright}
CERN-TH-2016-181
\end{flushright}

\vspace{1cm}
\begin{center}
{\grande Glauber theory and the quantum coherence}\\
\vspace{5mm}
{\grande  of curvature inhomogeneities}\\
\vspace{1cm}
 Massimo Giovannini 
 \footnote{Electronic address: massimo.giovannini@cern.ch} \\
\vspace{1cm}
{{\sl Department of Physics, 
Theory Division, CERN, 1211 Geneva 23, Switzerland }}\\
\vspace{0.5cm}
{{\sl INFN, Section of Milan-Bicocca, 20126 Milan, Italy}}
\vspace*{1cm}
\end{center}

\vskip 0.3cm
\centerline{\medio  Abstract}
\vskip 0.1cm
The curvature inhomogeneities are systematically scrutinized in the framework 
of the Glauber approach. The amplified quantum fluctuations of the 
scalar and tensor modes of the geometry are shown to be first-order coherent while 
the interference of the corresponding intensities is larger than in the case 
of Bose-Einstein correlations. After showing that the degree of second-order coherence 
does not suffice to characterize unambiguously the curvature inhomogeneities, 
we argue that direct analyses of the degrees of third and fourth-order coherence are necessary to discriminate 
between different correlated states and to infer more reliably the statistical properties of the large-scale fluctuations. 
We speculate that the moments of the multiplicity distributions of the relic phonons might be 
observationally accessible thanks to new generations of instruments able to count the single photons of the Cosmic Microwave Background in the THz region.

\noindent

\vspace{5mm}
\vfill
\newpage
\renewcommand{\theequation}{1.\arabic{equation}}
\setcounter{equation}{0}
\section{Quantum fluctuations and their coherence}
\label{sec1}
The Cosmic Microwave Background radiation (CMB in what follows) has been repeatedly scrutinized by a number of ground based observations and various spaceborne missions starting with the COBE satellite \cite{COBE}.  The first data released by the WMAP collaboration  \cite{WMAP1} 
used the newly discovered cross-correlations between the temperature and the polarization to analyze the initial conditions of the Einstein-Boltzmann hierarchy.  The subsequent releases of the WMAP experiment and the Planck explorer results \cite{WMAP2} confirmed (and partially refined) the early determinations of the first WMAP data \cite{WMAP1} so that today we can say, with a fair degree of confidence, that  the initial conditions of the CMB temperature and polarization anisotropies are predominantly adiabatic and Gaussian. Every deviation from this paradigm leads to a number of entropic solutions\footnote{The current data \cite{WMAP1,WMAP2,ACBARQUAD} seem to suggest that a small fraction of anticorrelated entropic modes in the presence of a dominant adiabatic mode may even improve the fit of the temperature autocorrelations accounting for potential large-scale suppressions of the corresponding angular power spectra. } (see e.g.\cite{hh}). When the CMB data are combined with other cosmological data sets (such as the large-scale structure data \cite{LSS} and the supernova data \cite{SNN}) the typical parameters describing the large-scale curvature modes can be slightly (but not crucially) modified.  

The temperature and the polarization anisotropies of the CMB are believed to originate in the early Universe from the fluctuations of the spatial curvature. Sakharov \cite{sakharov} was presumably the first one to raise the question of the quantum mechanical origin of density perturbations in the early Universe suggesting that the complicated patterns observed in the galaxy counts could actually have some plausible origin in the zero-point fluctuations of certain quantum fields.  The adiabatic and Gaussian nature of the initial conditions of the Einstein-Boltzmann hierarchy is compatible with a quantum mechanical origin of the curvature perturbations but the observed patterns of the temperature and polarization anisotropies are not sufficient 
to establish their quantum origin. The quest for accuracy in the determinations of the parameters of the
concordance paradigm does not tell us anything about the quantum state of large-scale perturbations and of their multiplicity
distribution\footnote{In what follows we shall often refer to the statistical properties of quantum states and of stochastic processes with discrete state space. By this we mean their degree of coherence encoding the moments of the multiplicity distribution. See, in this respect, sections \ref{sec2} and \ref{sec3}.}.

The quantum state of the scalar and tensor modes of the 
geometry is determined by the dynamical features of the model, by the its initial conditions and, at least in the standard lore, by the overall duration of the inflationary phase. Even if there are some who suggest that we have an accurate control of the protoinflationary dynamics
it would be nice to develop a set of sufficient criteria enabling us to infer the quantum origin of large-scale curvature perturbations from 
some  observational evidence. A first step along this direction relies on the idea of studying (and eventually measuring) the correlation functions of the intensities of the curvature perturbations \cite{max1} rather than the correlations of the corresponding amplitudes. This concept has been originally proposed by Hanbury Brown and  Twiss \cite{HBT0} and the analysis of the intensity correlations is often dubbed Hanbury Brown-Twiss (HBT) interferometry as opposed to the 
standard Young-type interference where only amplitudes (rather than intensities) are concerned\footnote{
The applications of the HBT  effect range from stellar astronomy \cite{HBT0}  to subatomic physics \cite{revs} where the interference of the intensities has been used to determine the hadron fireball dimensions \cite{cocconi} corresponding, in rough terms, to the linear size of the interaction region in proton-proton collisions. }.

The Glauber theory has been used through the years to reach a fuller understanding of the relations between the classical and quantum theories of light \cite{mandel}. The first experimental detection of non-classical light was made nearly fifty years ago \cite{burnham}. Since then the second-order coherence effects have been widely used to infer the statistical properties of light sources.  For similar purposes the Glauber theory can be applied to the analysis of the large-scale curvature fluctuations.  In the conventional view the questions raised by this approach are dismissed since the quantum origin of large scale curvature fluctuations is, according to the theoretical prejudice, an indisputable fact of nature.  In this paper we would like to take a pragmatic attitude which is incidentally also more modest:  we ought to understand in what sense the HBT interferometry \cite{max1} is sufficient to infer the statistical properties of large-scale curvature perturbations. Even if the theoretical results addressed here are encouraging, a number of experimental questions remain. While these questions are beyond the scopes of this investigation and will be left unanswered now, it is not excluded they might find satisfactory answers in the future. 

The layout of this investigation is therefore the following. In section \ref{sec2} we shall summarize the essentials of the Glauber 
approach by stressing the physical analogies with the treatment  of large-scale 
curvature perturbations. The first and second-order coherence effects will be analyzed in section \ref{sec3}.
The ambiguities of the second-order coherence will be addressed in section \ref{sec4}. In section \ref{sec5} we shall then argue 
that the degrees of third and fourth-order coherence are sufficient to infer the statistical 
properties of large-scale curvature inhomogeneities. In section \ref{sec6}  we shall discuss some future 
observational perspectives. Finally, section \ref{sec7}  contains our concluding remarks. To avoid lengthy digressions some of the formulas instrumental in obtaining the results presented in the bulk of the paper have been rederived, 
in a self-contained perspective, in the appendices \ref{APPA}, \ref{APPB} and \ref{APPC}.

\renewcommand{\theequation}{2.\arabic{equation}}
\setcounter{equation}{0}
\section{Glauber theory of coherence}
\label{sec2}
Consider the field operator $\hat{q}(\vec{x}, \tau)$  where $\vec{x}$ is the position and $\tau$ is the (conformal) time coordinate.
In quantum optics  $\hat{q}$ is often identified with a single polarization of the electric field (as in the original Glauber approach \cite{glauber}), with a single polarization of the vector potential (as argued by Mandel and Wolf\footnote{The authors of Refs. \cite{mandel} suggested indeed that the definition in terms of the vector potential is more convenient when dealing with the fotoelectric detection of light fluctuations.} \cite{mandel}) or even with a vector field itself, if suitable vector indices are included.

In the present context the field operator $\hat{q}$ coincides either with the canonically normalized curvature perturbations of a conformally flat\footnote{Consistently with observational determinations \cite{WMAP1,WMAP2},  the background metric will be conformally flat, i.e. $\overline{g}_{\mu\nu} = a^2(\tau) \eta_{\mu\nu}$ where $a(\tau)$ is the scale factor, $\tau$ denotes the conformal 
time coordinate and $\eta_{\mu\nu}$ is the flat metric with signature mostly minus, i.e. $(+, \,-,\, -,\,-)$.} Friedmann-Robertson-Walker space-time or with a single tensor polarization. In both cases the evolution of the quantum Hamiltonian can be written as: 
\begin{equation}
\hat{H}(\tau) = \frac{1}{2} \int d^{3} x \biggl[ \hat{\pi}^2 - 2\, i\,  \lambda ( \hat{\pi} \hat{q} + \hat{q} \hat{\pi}) + \partial_{k} \hat{q} \partial^{k} \hat{q}\biggr],
\label{AA1aa}
\end{equation}
where $\hat{q}$ and $\hat{\pi}$ are the two canonically conjugate and Hermitian field operators; the pump field $\lambda$ 
depends on the conformal time and it satisfies $\lambda^{*} = - \lambda$. According to the 
specific dictionary developed in the appendix \ref{APPA}, Eq. (\ref{AA1aa}) may describe the parametric amplification of either the scalar or of the tensor modes of the geometry. Equation (\ref{AA1aa}) is equivalent to Eq. (\ref{AA4})
which has been analyzed for the first time by Mollow and Glauber \cite{sq1} in the framework of the quantum theory of the parametric amplification. When expressed in terms of the corresponding creation and annihilation operators the free part of the Hamiltonian and the two components of the interacting Hamiltonian satisfy the usual commutation relations of the $SU(1,1)$ Lie algebra. 

In the case of the tensor modes of the geometry the canonical Hamiltonian of Eq. (\ref{AA1aa}) follows from the corresponding gauge-invariant action obtained long ago by Ford and Parker\footnote{In the seminal papers of Grishchuk \cite{gr0} the emphasis was on the breaking of Weyl invariance of the equations of motion rather than on the gauge-invariant action.} \cite{fordp}.  The scalar 
fluctuations of the metric will be described here within the standard Bardeen formalism \cite{bard1}; the corresponding gauge-invariant action has been discussed, for the first time, by Lukash \cite{luk} in a background model filled by a perfect and irrotational fluid. Various authors applied a similar analysis to the case of scalar field matter and to its fluctuations \cite{KS,chibisov}. 
The Hamiltonians for the scalar and tensor modes have been discussed, in a unified perspective, in appendix \ref{APPA} (see also \cite{mg1}).  
Indeed, the same process leading to the curvature inhomogeneities also produces to a stochastic background of relic gravitons extending over three decades in frequency \cite{maxg1}.

 \subsection{The Glauber correlation function}
The parametric amplification described by the Hamiltonian (\ref{AA1aa}) produces specific quantum correlations in the final state.
Since the degrees of quantum coherence provide a systematic approach to the statistical properties of the final state \cite{glauber},
it is rather plausible to analyze the large-scale curvature perturbations in the light of the Glauber approach.
To comply with this program, the first step is to introduce the Glauber correlation function in its most general form:
\begin{equation}
{\mathcal G}^{(n,m)}(x_{1}, \,.\,.\,.\,x_{n}, \, x_{n+1},\, .\,.\,.\,, x_{n +m}) = \mathrm{Tr}\biggl[ \hat{\rho} \, \hat{q}^{(-)}(x_{1})\,.\,.\,.\, \hat{q}^{(-)}(x_{n})
\, \hat{q}^{(+)}(x_{n+1})\,.\,.\,.\, \hat{q}^{(+)}(x_{n+m})\biggr],
\label{corr1}
\end{equation}
where $x_{i} \equiv (\vec{x}_{i}, \, \tau_{i})$ and $\hat{\rho}$ is the density operator representing the (generally mixed) state of the field $\hat{q}$.  The field $\hat{q}(\vec{x}, \tau)$ can always be expressed as $\hat{q}(x) = \hat{q}^{(+)}(x) + \hat{q}^{(-)}(x)$, with $\hat{q}^{(+)}(x)= \hat{q}^{(-)\,\dagger}(x)$.
By definition we will have that $\hat{q}^{(+)}(x) |\mathrm{vac} \rangle=0$ and also that  $\langle \mathrm{vac} |\, \hat{q}^{(-)}(x) =0$;  the state $|\mathrm{vac}\rangle $ denotes the vacuum\footnote{ 
In the discussion of section \ref{sec3} the vacuum corresponds to the state minimizing the Hamiltonian at the onset of the dynamical evolution as discussed in appendix \ref{APPB}. This state can be  explicitly constructed by diagonalizing the Hamiltonian in terms of an appropriate canonical 
transformation. A similar procedure is used to derive the ground state wavefunction of an interacting Bose gas at zero 
temperature \cite{fetter,solomon}.}. Provided the total duration of inflation exceeds the minimal number of about $65$ efolds \cite{WMAP2,mg1,liddle}, the vacuum initial data are the most plausible, at least in the conventional lore.

Let us now consider in greater detail the expression of Eq. (\ref{corr1}) and let us observe that an operator of the type  
\begin{equation}
\hat{O}(x_{1}, \,.\,.\,.\,x_{n}) = \hat{q}^{(-)}(x_{1})\,.\,.\,.\, \hat{q}^{(-)}(x_{n})
\, \hat{q}^{(+)}(x_{1})\,.\,.\,.\, \hat{q}^{(+)}(x_{n}),
\label{corr3}
\end{equation}
is needed to describe $n$-fold delayed coincidence measurements of the field at the space-time points $(x_{1}, \,.\,.\,.\,x_{n})$.
If $|\, b\rangle$ is the state before the measurement and $|\, a\rangle$ is the state after the measurement, the matrix element corresponding 
to the absorption of the quanta of $\hat{q}$  at each detector and at given times is $\langle a\, | \hat{q}^{(+)}(x_{1})\,.\,.\,.\, \hat{q}^{(+)}(x_{n})|\, b\rangle$. The rate at which such absorptions occur, summed over the final states, is therefore proportional to: 
\begin{eqnarray}
&& \sum_{a} \biggl|\langle a\, | \hat{q}^{(+)}(x_{1})\,.\,.\,.\, \hat{q}^{(+)}(x_{n})|\, b\rangle \biggr|^2 = 
\nonumber\\
&& \sum_{a} \langle b| \hat{q}^{(-)}(x_{1})\,.\,.\,.\, \hat{q}^{(-)}(x_{n})| a\rangle \langle a| 
 \hat{q}^{(+)}(x_{1})\,.\,.\,.\, \hat{q}^{(+)}(x_{n}) |b \rangle = \langle b| \hat{O} |b \rangle,
\label{corr4}
\end{eqnarray}
where $\hat{O}$ has been given in Eq. (\ref{corr3}) and the second equality of Eq. (\ref{corr4}) follows from the completeness relation. It is clear from Eq. (\ref{corr4}) that when $\langle b| \hat{O} |b \rangle$ is averaged over the ensemble of the initial states of the system it becomes identical with Eq. (\ref{corr1}) for $x_{n + r} = x_{r}$ (with $r= 1, 2, \,.\,.\,., n$ and $n=m$). Since this is the case that will be studied hereunder we shall denote the Glauber correlation function as
\begin{equation}
{\mathcal G}^{(n)}(x_{1}, \,.\,.\,.\,x_{n}, \, x_{n+1},\, .\,.\,.\,, x_{2n}) = \mathrm{Tr}\biggl[ \hat{\rho} \, \hat{q}^{(-)}(x_{1})\,.\,.\,.\, \hat{q}^{(-)}(x_{n})
\, \hat{q}^{(+)}(x_{n+1})\,.\,.\,.\, \hat{q}^{(+)}(x_{2n})\biggr].
\label{corr5}
\end{equation}

\subsection{The degrees of quantum coherence}

Recalling the result of Eq. (\ref{corr5}), the coherence properties of the quantum field $\hat{q}(x)$ can be discussed by introducing the normalized version of  the $n$-point Glauber function \cite{glauber}:
\begin{equation}
g^{(n)}(x_{1}, \,.\,.\,.\,x_{n}, \, x_{n+1},\, .\,.\,.\,, x_{2n}) = \frac{{\mathcal G}^{(n)}(x_{1}, \,.\,.\,.\,x_{n}, \, x_{n+1},\, .\,.\,.\,, x_{2n})}{\sqrt{\Pi_{j =1}^{2 n}\, {\mathcal G}^{(1)}(x_{j}, x_{j})}}.
\label{corr6}
\end{equation}
While, by definition, $|g^{(1)}(x_{1},\, x_{2})| \leq 1$ the higher order correlators are not restricted in absolute value as it happens for $g^{(1)}(x_{1}, x_{2})$. A fully  coherent field  must therefore satisfy the following necessary condition:
\begin{equation}
g^{(n)}(x_{1}, \,.\,.\,.\,x_{n}, \, x_{n+1},\, .\,.\,.\,, x_{2n}) = 1,
\label{corr6a}
\end{equation}
for $ n = 1,\, 2,\, 3,\, .\,.\,.$.  If only a limited number of normalized correlation functions will satisfy Eq. (\ref{corr6a}) we shall speak about partial 
coherence. So for instance if $g^{(1)}(x_{1},\, x_{2})= 1$ and $g^{(2)}(x_{1},\, x_{2},\, x_{3},\, x_{4}) =1$ 
(but $g^{(3)}(x_{1},\, x_{2},\, x_{3},\, x_{4},\, x_{5},\,x_{6}) \neq 1$) we shall say that the radiation is second-order coherent. 
We shall be specifically interested in the first, second, third and fourth-order correlators. 
The degrees of first- and second-order coherence are:
\begin{eqnarray}
&& g^{(1)}(x_{1}, x_{2}) = \frac{{\mathcal G}^{(1)}(x_{1}, x_{2})}{\sqrt{{\mathcal G}^{(1)}(x_{1}, x_{1})\, {\mathcal G}^{(1)}(x_{2}, x_{2})}},
\label{g1}\\
&& g^{(2)}(x_{1}, x_{2}, x_{3}, x_{4}) = \frac{{\mathcal G}^{(2)}(x_{1}, x_{2}, x_{3}, x_{4})}{\sqrt{{\mathcal G}^{(1)}(x_{1}, x_{1})\, {\mathcal G}^{(1)}(x_{2}, x_{2})\, {\mathcal G}^{(1)}(x_{3}, x_{3})\,{\mathcal G}^{(1)}(x_{4}, x_{4})}}.
\label{g2}
\end{eqnarray}
Similarly the third- and fourth-order degrees of coherence are:
\begin{eqnarray}
g^{(3)}(x_{1},\,x_{2},\, x_{3},\, x_{4},\,x_{5},\, x_{6}) &=&  \frac{{\mathcal G}^{(3)}(x_{1},\,x_{2},\, x_{3},\, x_{4},\,x_{5},\,x_{6})}{\sqrt{\prod_{i =1}^{6} {\mathcal G}^{(1)}(x_{i},\, x_{i})}},
\label{g3}\\
g^{(4)}(x_{1},\, x_{2},\, x_{3},\, x_{4},\,x_{5}, x_{6},\, x_{7},\,x_{8}) &=&  \frac{{\mathcal G}^{(4)}(x_{1},\,x_{2},\, x_{3},\, x_{4},\,x_{5},\,x_{6},\,x_{7},\,x_{8})}{\sqrt{\prod_{i \,=\,1}^{8} {\mathcal G}^{(1)}(x_{i},\, x_{i})}}.
\label{g4}
\end{eqnarray}
The degree of first-order coherence appears naturally in the Young two-slit experiment. Whenever the degree of first-order coherence is equal to $1$ the visibility is maximized \cite{mandel}.
The degree of second-order coherence enters the discussion of the Hanbury Brown-Twiss effect \cite{HBT0} and its different applications 
ranging from stellar interferometry \cite{mandel} to high-energy physics \cite{revs,cocconi}. The degree of second-order coherence 
arises naturally when discussing the correlations of the intensities of the field $\hat{q}$ (see section \ref{sec3}). Notice that 
the intensity correlators relevant to the HBT interferometry can be easily obtained from Eq. (\ref{corr6}) by identifying the space-time points as 
follows:
\begin{equation}
x_{1} \equiv x_{n+1},\qquad x_{2} \equiv x_{n+2},\qquad .\,.\,.\, \qquad x_{n} \equiv x_{2 n}.
\label{ident}
\end{equation}
In this case the original Glauber correlator will effectively be a function of $n$ points and and it will describe the correlation 
of $n$ intensities $\hat{{\mathcal I}}(x_{i}) = \hat{q}(x_{i}) \hat{q}(x_{i})$ where $x_{i} = x_{1},\, x_{2}, \,.\,.\,. x_{n}$.

Equations (\ref{g3}) and (\ref{g4}) describe, respectively, the degrees of third-order and fourth-order coherence. It has been recently suggested, in quantum optical applications, that  the degree of second-order coherence might not always be  sufficient to specify completely the statistical properties of the radiation field \cite{opt1,opt2,opt3,opt4}. A specific discussion of these points can be found in section \ref{sec4}.

Since a fully coherent field must satisfy Eq. (\ref{corr6a}) the Poissonian case will be used as a benchmark value to analyze 
the higher-order correlators. From the definitions given above it can be argued that 
while the degree of first-order coherence cannot exceed $1$ (see Eq. (\ref{corr6a}) and discussion thereafter), the degrees of higher-order coherence do not have an obvious upper bound. 
Various inequalities satisfied by degrees of coherence of various orders have been derived by Titulaer and Glauber \cite{titulaer} but the simpler results given above are adequate, at least for the present discussion. 
\renewcommand{\theequation}{3.\arabic{equation}}
\setcounter{equation}{0}
\section{Super-Poissonian degree of coherence}
\label{sec3}
The degrees of first-order and second-order coherence given Eqs. (\ref{g1}) and (\ref{g2}) will now be specifically analyzed when the 
evolution of the field operators is governed by the Hamiltonian  (\ref{AA1aa}) and the corresponding pump fields
are determined by the quasi-de Sitter evolution. The initial conditions imposed on the evolution of the field operators will imply that the Hamiltonian of Eq. (\ref{AA1aa}) is both diagonal and minimized. Equation (\ref{corr5}) in the cases $n=1$ and $n=2$ implies:
\begin{eqnarray}
{\mathcal G}^{(1)}(x_{1},\, x_{2}) &=& \langle \hat{q}^{(-)}(x_{1}) \, \hat{q}^{(+)}(x_{2}) \rangle,
\label{CC1}\\
{\mathcal G}^{(2)}(x_{1},\, x_{2},\, x_{3},\, x_{4})  &=& \langle \hat{q}^{(-)}(x_{1}) \, \hat{q}^{(-)}(x_{2}) \hat{q}^{(+)}(x_{3}) \, \hat{q}^{(+)}(x_{4})\rangle.
\label{CC2}
\end{eqnarray}
The correlators ${\mathcal G}^{(1)}(x_{1},\, x_{2})$ and  ${\mathcal G}^{(2)}(x_{1},\, x_{2},\, x_{3},\, x_{4})$ obey the following relations:
\begin{equation}
{\mathcal G}^{(1)}(x_{1},\, x_{2}) = {\mathcal G}^{(1)\,*}(x_{2}, x_{1}), \qquad {\mathcal G}^{(2)}(x_{1},\, x_{2},\, x_{3},\, x_{4})  = 
{\mathcal G}^{(2)\,*}(x_{4},\, x_{3},\, x_{2},\, x_{1}).
\end{equation}
The quantum theory of the parametric amplification will be discussed in the Heisenberg 
description\footnote{The quantum theory of parametric amplification has been originally developed in \cite{sq1} and subsequently analyzed in connection with the dynamics of the squeezed quantum states \cite{sq2,sq3}. An explicit discussion of the evolution of cosmological perturbations in the Schr\"odinger representation has been presented long ago by Grishchuk and collaborators Ref. \cite{gr1} and subsequently discussed by various authors (see for instance \cite{max1,mg2}); see also the discussion at 
 the end of appendix \ref{APPC}.}. When the pump field $\lambda$ is determined by the quasi-de Sitter the degree of first-order coherence goes always to $1$; 
this happens in the limit of wavelengths larger than the Hubble radius and in spite of the correlation properties of the initial state. In the same limits the degree of second-order coherence goes to $3$; the curvature phonons (and the gravitons) are therefore bunched \cite{mandel,loudon}, their statistics is super-Poissonian and their degree of correlation exceeds the typical value of a chaotic source.

\subsection{Quasi-de Sitter evolution: first-order coherence}
The evolution of the field operators depends on two complex functions $v_{k}(\tau)$ and $u_{k}(\tau)$ whose 
explicit expression can be obtained by solving Eqs. (\ref{AA14}) and (\ref{AA15}). In the case of a quasi-de Sitter evolution 
$v_{k}(\tau)$ and $u_{k}(\tau)$ are given by Eqs. (\ref{AA18}) and (\ref{AA18a}); n the pure de Sitter case their expression 
is given instead by Eq. (\ref{AA19}).  The first-order correlation function of Eq. (\ref{CC1}) at separate space-time points:
 \begin{equation}
{\mathcal G}^{(1)}(x_{1},\, x_{2}) = \frac{1}{2 (2\pi)^3} \int \frac{d^{3}k_{1}}{k_{1}} \, v^{*}_{k_{1}}(\tau_{1}) \, v_{k_{1}}(\tau_{2}) \, e^{ - i \vec{k}_{1}\cdot (\vec{x}_{1} - \vec{x}_{2}) }.
\label{CC3}
\end{equation}
Consequently, from Eqs. (\ref{g1}) and (\ref{CC3})  the normalized degree of first-order coherence is:
\begin{equation}
g^{(1)}(\vec{x}_{1},\, \vec{x}_{2};\, \tau_{1},\, \tau_{2}) = \frac{\int d^{3} k_{1} \, v_{k_{1}}^{*}(\tau_{1})\, v_{k_{1}}(\tau_{2})/k_{1} \, e^{- i \vec{k}_{1} \cdot(\vec{x}_{1} - \vec{x}_2)}}{\sqrt{\int d^{3} k_{1} \, v_{k_{1}}^{*}(\tau_{1})\, v_{k_{1}}(\tau_{2})/k_{1}} \sqrt{\int d^{3} k_{2} \, v_{k_{2}}^{*}(\tau_{2})\, v_{k_{2}}(\tau_{2})/k_{2}}}.
\label{CC4}
\end{equation}
We are interested in the value of the first-order coherence when the relevant wavelengths are larger than the Hubble radius and, in this 
limit, the explicit expressions of Eqs. (\ref{AA18}) and (\ref{AA18a}) imply:
\begin{equation}
v_{k}(\tau) = - i \,\frac{\Gamma(\mu)}{\sqrt{2}\pi} \, e^{- i\pi(\mu + 1/2)/2}\, \biggl( - \frac{k \tau}{2} \biggr)^{1/2 - \mu} \biggl[ 1 + {\mathcal O}(3- 2\mu)\biggr] + 
{\mathcal O}(|k\tau|^{1 -\mu}),
\label{CC4a}
\end{equation}
where $\mu$ is given, in the scalar and tensor case, by Eq. (\ref{AA18a}). 
Equation (\ref{CC4a}) has been obtained by expanding the Hankel functions of Eq. (\ref{AA18}) in the limit of small arguments \cite{abr}.
The first subleading term appearing in Eq. (\ref{CC4a}) is suppressed by $3 - 2 \mu$ while the second correction is suppressed, in comparison with the leading term, by $|k\tau|^{1/2}$ which is small in the large-scale 
limit\footnote{In the scalar case $ 3 - 2\mu_{scalar} = (2 \eta -\epsilon)/(1 -\epsilon)$ while in the tensor case $ 3 - 2\mu_{tensor} = -\epsilon/(1 -\epsilon)$: in both situations 
these combinations are suppressed by the slow-roll parameters $\epsilon = - \dot{H}/H^2$ and $\eta = \ddot{\varphi}/( H \dot{\varphi})$ where 
$H$ and $\varphi$ are, respectively, the Hubble rate during inflation and the inflaton field (see also Eq. (\ref{AA18a})).}. In the limit $\mu \to 3/2$ we have, from Eq. (\ref{CC4a}), that 
$v_{k}(\tau) = - i/(2 k\tau)$ which coincides with the exact result obtainable by setting $\mu=3/2$ in Eq. (\ref{AA18}). 
Performing the angular integrals we have that Eq. (\ref{CC4a}) implies 
\begin{equation}
g^{(1)}(\vec{r};\, \tau_{1},\, \tau_{2}) = \frac{\int k_{1} d k_{1} \, v_{k_{1}}^{*}(\tau_{1})\, v_{k_{1}}(\tau_{2})\, j_{0}(k_{1} r)}{\sqrt{\int k_{1} d k_{1} \, v_{k_{1}}^{*}(\tau_{1})\, v_{k_{1}}(\tau_{2})}\sqrt{\int k_{1} d k_{1} \, v_{k_{1}}^{*}(\tau_{1})\, v_{k_{1}}(\tau_{2})}},
\label{CC5}
\end{equation}
where $j_{0}(k_{1} r)$ denotes the spherical Bessel function of zeroth order \cite{abr}. The numerator and the denominator of Eq. (\ref{CC5}) 
depend on $\tau_{1}$ and $\tau_{2}$. However, owing to the specific form of $v_{k}(\tau)$, the dependence on $\tau_{1}$ and $\tau_{2}$ 
simplifies and the final form of Eq. (\ref{CC5}) becomes
\begin{equation}
g^{(1)}(r) = \frac{\int d k_{1}\, k^{ 2 - 2 \mu} \, j_{0}(k r_{1})}{\int d k_{1}\, k^{ 2 - 2 \mu}} \to 1,
\label{CC5a}
\end{equation}
where the integrals are evaluated over all the modes larger than the Hubble radius and the second relation clearly holds in the 
limit $k_{1} r \ll 1$ (corresponding to large angular separations). The same result implied by Eq. (\ref{CC5a}) can be obtained when 
$ k\tau \gg 1$ in Eq. (\ref{AA18}). In this case the Hankel functions become plane waves and 
in the limit $k_{1} r \ll 1$ we have that $g^{(1)}(\vec{r},\tau_{1},\tau_{2})$ always tend to $1$ implying the first-order coherence of the underlying fluctuations. 

\subsection{Quasi-de Sitter evolution: second-order coherence}
The general form of the second-order correlation function involves four separated space-time points. As in the case of the first-order coherence Eq. (\ref{CC2}) can be written in terms of $v_{k}(\tau)$ and $u_{k}(\tau)$:
\begin{eqnarray}
{\mathcal G}^{(2)}(x_{1},\, x_{2},\, x_{3},\, x_{4}) &=& \int \frac{d^{3} k_{1}}{2 k_{1} (2\pi)^3} \int \frac{d^{3} k_{2}}{2 k_{2} (2\pi)^3} 
\nonumber\\
&\times& \biggl[ v_{k_{1}}^{*}(\tau_1)\,v_{k_{2}}^{*}(\tau_2) \,v_{k_{1}}(\tau_{3})\, v_{k_{2}}(\tau_{4}) e^{- i \vec{k}_{1}\cdot(\vec{x}_{1} - \vec{x}_{3})} e^{- i \vec{k}_{2}\cdot(\vec{x}_{2} - \vec{x}_{4})}
\nonumber\\
&+& v_{k_{1}}^{*}(\tau_1)\,v_{k_{2}}^{*}(\tau_2) \,v_{k_{1}}(\tau_{4})\, v_{k_{2}}(\tau_{3}) e^{- i \vec{k}_{1}\cdot(\vec{x}_{1} - \vec{x}_{4})} e^{- i \vec{k}_{2}\cdot(\vec{x}_{2} - \vec{x}_{3})}
\nonumber\\
&+& v_{k_{1}}^{*}(\tau_1)\,u_{k_{1}}^{*}(\tau_2) \,u_{k_{2}}(\tau_{3})\, v_{k_{2}}(\tau_{4}) e^{- i \vec{k}_{1}\cdot(\vec{x}_{1} - \vec{x}_{2})} e^{- i \vec{k}_{2}\cdot(\vec{x}_{3} - \vec{x}_{4})} \biggr].
\label{CC6}
\end{eqnarray}
If we identify the space-time points two by two (i.e $x_{1}= x_{3}$ and $x_{2} = x_{4}$) the correlation function of Eq. (\ref{CC6}) 
describes the interference of two beams with intensities $\hat{{\mathcal I}}(\vec{x}_{1}, \tau_{1})$ and $\hat{{\mathcal I}}(\vec{x}_{2}, \tau_{2})$, i.e. 
\begin{eqnarray}
{\mathcal G}^{(2)}(x_{1},\, x_{2})  &=& \langle \hat{{\mathcal I}}(\vec{x}_{1}, \tau_{1}) \, \hat{{\mathcal I}}(\vec{x}_{2}, \tau_{2}) \rangle =
\int \frac{d^{3} k_{1}}{2 k_{1} (2\pi)^3} \int \frac{d^{3} k_{2}}{2 k_{2} (2\pi)^3} 
\nonumber\\
&\times& \biggl\{ |v_{k_{1}}(\tau_{1})|^2 \, |v_{k_{2}}(\tau_{2})|^2 \biggl[ 1 + e^{- i (\vec{k}_{1}- \vec{k}_{2}) \cdot\vec{r}}\biggr]
\nonumber\\
&+& v_{k_{1}}^{*}(\tau_1)\,u_{k_{1}}^{*}(\tau_2) \,u_{k_{2}}(\tau_{1})\, v_{k_{2}}(\tau_{2})\,e^{- i (\vec{k}_{1}+ \vec{k}_{2}) \cdot\vec{r}}\biggr\},
\label{CC6a}
\end{eqnarray}
where $\vec{r} = \vec{x}_{1} - \vec{x}_{2}$. Inserting Eqs. (\ref{CC6}) and (\ref{CC6a})  into Eq. (\ref{g2})  the degree of second-order coherence 
becomes:
\begin{eqnarray}
g^{(2)}(\vec{r}, \tau_{1},\tau_{2}) &=& \frac{ \langle \hat{{\mathcal I}}(\vec{x}_{1}, \tau_{1}) \, \hat{{\mathcal I}}(\vec{x}_{2}, \tau_{2})\rangle}{\langle \hat{{\mathcal I}}(\vec{x}_{1}, \tau_{1})\rangle \langle  \hat{{\mathcal I}}(\vec{x}_{2}, \tau_{2}) \rangle} 
\nonumber\\
&=& 1 + \frac{\int k_{1} d k_{1} |v_{k_{1}}(\tau_{1})|^2 \, j_{0}(k_{1} r) \,\,\int k_{2} d k_{2} |v_{k_{2}}(\tau_{2})|^2\, j_{0}(k_{2} r) }{\int k_{1} \,d k_{1} |v_{k_{1}}(\tau_{1})|^2\,\int k_{2} \,d k_{2} |v_{k_{2}}(\tau_{2})|^2}
\nonumber\\
&+&  \frac{\int k_{1} d k_{1} \,u_{k_{1}}^{*}(\tau_{2}) v_{k_1}^{*}(\tau_{1})\, j_{0}(k_{1} r) \,\,\int k_{2} d k_{2} \, u_{k_2}(\tau_{1})v_{k_{2}}(\tau_{2})\,j_{0}(k_{2} r)}{\int k_{1} d k_{1} |v_{k_{1}}(\tau_{1})|^2 \,\,\int k_{2} d k_{2} |v_{k_{2}}(\tau_{2})|^2}.
\label{CC8}
\end{eqnarray}
According to Eq. (\ref{CC8}) the degree of second-order coherence involves four field operators at two separated spatial points \cite{mandel,loudon}. 
From the observational viewpoint the correlations of two intensities (i.e. two beams) translates into the correlation of the output of four distinct brightness perturbations (see also the discussion in section \ref{sec6}). Equation (\ref{CC8}) can be further simplified by appreciating that the mean number of produced particles per Fourier mode is simply $\overline{n}_{k_{1}}(\tau_{1}) = |v_{k_{1}}(\tau_{1})|^2$ and $\overline{n}_{k_{2}}(\tau_{2}) = |v_{k_{2}}(\tau_{2})|^2$. Furthermore in the pure de Sitter case Eq. (\ref{AA18a}) imply:
\begin{eqnarray}
\frac{ u_{k_{1}}^{*}(\tau_{1}) \, v_{k_{1}}^{*}(\tau_{1})  u_{k_{2}}(\tau_{2}) \, v_{k_{2}}(\tau_{2}) }{|v_{k_{1}}(\tau_{1})|^2 \,  |v_{k_{2}}(\tau_{2})|^2}=  
 1+ 4 k_{1} \tau_{1} k_{2} \tau_{2} + 2 i\, \tau(k_{2}\tau_{2} - k_{1}\tau_{1}).
\label{CC9}
\end{eqnarray}
The same relation holds also in the quasi-de Sitter case to leading order in the slow-roll expansion and for scales 
larger than the Hubble radius (i.e. $k\tau \ll 1$). Thanks to the preceding two observations we have that the degree of second-order coherence,
to leading order\footnote{The normal-ordered expectation values dominate the degrees of coherence since the average multiplicity of the produced phonons and gravitons is large; this observation defines the $1/\overline{n}_{k}$ expansion (see, in this respect, the final part of appendix \ref{APPC}). In the pure de Sitter case, from Eq. (\ref{AA19}),  $\overline{n}_{k}(\tau) =
(4 k^2 \tau^2)^{-1}\gg 1$ for $k\tau\ll 1$.} in $1/\overline{n}_{k}$ is given by:
\begin{eqnarray}
g^{(2)}(\vec{r},\tau_{1},\tau_{2}) &=& 1 + 2 \frac{\int k_{1} d k_{1} j_{0}(k_{1} \, r)\, \overline{n}_{k_{1}}(\tau_{1}) \, \int k_{2} d k_{2} j_{0}(k_{2} \, r)\, \overline{n}_{k_{2}}(\tau_{2})}{\int k_{1} d k_{1}  \overline{n}_{k_{1}}(\tau_{1}) \, \int k_{2} d k_{2}\, \overline{n}_{k_{2}}(\tau_{2})}
\nonumber\\
&+&  \frac{\int k_{1} d k_{1} j_{0}(k_{1} \, r)/\sqrt{\overline{n}_{k_{1}}(\tau_{1})}\,  \, \int k_{2} d k_{2} j_{0}(k_{2} \, r)/\sqrt{\overline{n}_{k_{2}}(\tau_{2})}\, }{\int k_{1} d k_{1} \overline{n}_{k_{1}}(\tau_{1})\int k_{2} d k_{2}\, \overline{n}_{k_{2}}(\tau_{2})}.
\label{CC10}
\end{eqnarray}
The large-scale limit the spherical Bessel functions go to $1$ and therefore 
Eq. (\ref{CC10}) becomes\footnote{The result of Eq. (\ref{CC11}) holds also in the case $\tau_{1} \to \tau_{2}$;  in the zero time-delay limit 
$\tau_{1}- \tau_{2}=0$ and in this case the degree of second-order coherence will simply depend on $\tau$ and $\vec{r}$.}:
\begin{equation}
g^{(2)}(r,\tau_{1}, \tau_{2}) \to  3, \qquad \lim_{\tau_{1} \to \tau_{2}} g^{(2)}(r,\tau_{1}, \tau_{2}) = g^{(2)}(r,\tau).
\label{CC11}
\end{equation}
The result of Eq. (\ref{CC11}) holds provided the total number of efolds $N_{tot}$ is larger than the maximal number of efolds 
today accessible by observations\footnote{The value of $N_{max}$ depends in turn on the post-inflationary history and conservative 
estimates suggest $N_{max} =  63 \pm 15$ \cite{max1,liddle}. If the post-inflationary history is standard (with sudden reheating) 
$N_{max} = 63.6 + \frac{1}{4} \ln{\epsilon}$. In this case $N_{max}$ coincides approximately with the minimal number of efolds $N_{\mathrm{min}}$ needed to solve the kinematic problems of the standard cosmological model. All in all the result of Eq. (\ref{CC11}) is pretty robust 
except when $N_{tot} \sim N_{max} \sim N_{min}$. In this case the correlations of the initial state may affect the degree of quantum coherence.
There are some who think that for the consistency  of the inflationary scenarios we must anyway demand that $N_{tot}$ exceeds $N_{max}$.}
denoted by $N_{max}$. The ambition of the present analysis is not to endorse a particular initial state 
of the curvature inhomogeneities but rather to stress the need of a systematic 
statistical scrutiny of the large-scale correlations. Given the larger arbitrariness of the initial data,  the
analysis of the degrees of higher-order coherence seems even more compelling when the initial state is not the vacuum.

The methods employed in subatomic physics involve the construction of two-particle (and eventually three-particle) correlation functions from the distribution of particles radiated from a hot (and spatially localized) source \cite{revs}. While the concepts are very much the same, the notations may be slightly different so that the pivotal variable is often defined as $R(\vec{r},\tau)= g^{(2)}(\vec{r},\tau) -1$. 

All in all the large-scale curvature fluctuations are only partially coherent. They are first-order coherent but, according to the Glauber terminology,  their statistics is  super-Poissonian with a degree of second-order coherence which is thrice the one of a fully coherent 
field (see Eq. (\ref{corr6a}) and discussion thereafter).

\renewcommand{\theequation}{4.\arabic{equation}}
\setcounter{equation}{0}
\section{Potential ambiguities}
\label{sec4}
In conventional HBT interferometry the space-time dimensions of the emitters need to be determined but the statistical properties of the source are known. For large-scale curvature perturbations the reverse is true and the gross uniformity of the temperature fluctuations at last scattering implies that curvature perturbations, prior to matter-radiation equality, had typical wavelengths larger than the Hubble radius at the corresponding epoch. In this situation the degrees of first and second-order coherence are however not sufficient to disambiguate the statistical properties of large-scale curvature inhomogeneities. This aspect can be already appreciated by noticing that the degree of second-order coherence 
of large-scale curvature inhomogeneities turns out to be numerically very close to the case of Bose-Einstein correlations.

\subsection{Bose-Einstein correlations}

Indeed the result of Eq. (\ref{CC11}) can be usefully compared with the one of Bose-Einstein correlations in the same large-scale limit. 
Let us then suppose, for the sake of comparison, that the field $\hat{q}(\vec{x}, \tau)$ is in a thermal state in standard 
Minkowski space-time. In this case the density matrix can be written, in the Fock basis, as: 
\begin{equation}
\hat{\rho} = \sum_{\{n\}} \, P_{\{n\}} \, | \{n\} \rangle \langle \{n\}|,\qquad \sum_{\{n\}} \, P_{\{n\}} =1.
\label{AMB1}
\end{equation}
The multimode probability distribution appearing in Eq. (\ref{AMB1}) is given by:
\begin{equation}
P_{\{n\}} = \prod_{\vec{k}} \frac{\overline{n}_{k}^{n_{\vec{k}}}}{( 1 + \overline{n}_{k})^{n_{\vec{k}} + 1 }},
\label{AMB2}
\end{equation}
where $\overline{n}_{k} = \mathrm{Tr}[ \hat{\rho} \, \hat{d}_{\vec{k}}^{\dagger}\, \hat{d}_{\vec{k}}]$ is the average occupation number of each Fourier mode. Furthermore, following the standard notation, $ |\{n \}\rangle = |n_{\vec{k}_{1}} \rangle \, | n_{\vec{k}_{2}} \rangle \, | n_{\vec{k}_{3}} \rangle...$ 
where the ellipses stand for all the occupied modes of the field. 
Even though the density matrix describes a mixed state, the average multiplicity in each Fourier mode(i.e.  $\overline{n}_{k}$)  does not need to coincide with the Bose-Einstein occupation number\footnote{ This situation occurs in quantum optics for chaotic (i.e. white) light 
where photons are distributed as in Eq. (\ref{AMB2}) for each mode of the radiation field but they
are produced by sources in which atoms are kept at an excitation level higher than that in thermal equilibrium \cite{loudon}.}.  Using Eq. (\ref{g2}) and performing the appropriate averages in 
by means of the density matrix of Eq. (\ref{AMB1}) the normalized degree of second-order coherence becomes, in this case,
\begin{equation}
g^{(2)}(\vec{r}, \tau_{1}, \tau_{2}) = \frac{\int d^{3} k_{1} \overline{n}_{k_{1}}(\tau_{1})/k_{1} \, \int d^{3} k_{2} \,\overline{n}_{k_{2}}/k_{2}\, \biggl[ 1 + e^{- i (\vec{k}_{1} + \vec{k}_{2})\cdot \vec{r}}\biggr]}{\int d^{3} k_{1} \overline{n}_{k_{1}}(\tau_{1})/k_{1} \, \int d^{3} k_{2} \,\overline{n}_{k_{2}}(\tau_{2})/k_{2}}.
\label{AMB3}
\end{equation}
While the calculation leading to Eq. (\ref{AMB3}) is a bit lengthy we just recall that the expectation value contains creation and annihilation operators with four different momenta; three typical leading terms will arise:  in the first term all the momenta coincide while in the two remaining terms the momenta are paired two by two. The expectation value becomes\footnote{All the momenta are equal  in the first line of Eq. (\ref{AMB4}); in the 
second line of Eq. (\ref{AMB4}) the momenta are paired two by two is such a way that double counting is avoided (see e.g. \cite{delcampo}).}
\begin{eqnarray}
&& \langle \hat{d}_{i}^{\dagger} \,  \hat{d}_{j}^{\dagger} \, \hat{d}_{k}\,  \hat{d}_{\ell} \rangle = 
\langle \hat{d}_{i}^{\dagger} \,  \hat{d}_{i}^{\dagger} \, \hat{d}_{i}\,  \hat{d}_{i}\rangle \delta_{i\,j} \,
\delta_{j\,k}\, \delta_{\ell\,k} +
\nonumber\\
&& \langle \hat{d}_{i}^{\dagger} \,  \hat{d}_{j}^{\dagger} \, \hat{d}_{i}\,  \hat{d}_{j} \rangle  \, \delta_{i\, k} \,
 \delta_{j\, \ell} [ 1 - \delta_{ij}] +  \langle \hat{d}_{i}^{\dagger} \,  \hat{d}_{j}^{\dagger} \, \hat{d}_{j}\,  \hat{d}_{i} \rangle  \, \delta_{i\, \ell} \,
 \delta_{j\, k} [ 1 - \delta_{ij}], 
\label{AMB4}
\end{eqnarray}
where $\hat{d}_{i}$ and $\hat{d}_{j}^{\dagger}$ denote the annihilation and creation operators related two generic momenta, 
i.e. for instance $\hat{d}_{\vec{q}}$ and $\hat{d}^{\dagger}_{\vec{p}}$; furthermore, following the same shorthand 
notation, $\delta_{i\, j}$ denotes the delta functions over the three-momenta (i.e. 
$ \delta_{\vec{q},\, \vec{p}}$).  Performing each of the averages of Eq. (\ref{AMB4}) in terms of the density matrix (\ref{AMB1}) 
leads to Eq. (\ref{AMB3}) after careful use of the various delta functions over the momenta.
We can finally integrate over the angular coordinates and the result will be\footnote{As in the case of Eq. (\ref{CC11}) the result of Eq. (\ref{AMB5a}) 
holds also in the in the zero time-delay limit  $\tau_{1} \to \tau_{2}$.}: 
\begin{equation}
g^{(2)}(\vec{r}, \tau) = 1 + \frac{\int k_{1} d k_{1} \overline{n}_{k_{1}}(\tau)\,j_{0}(k_{1} r)  \int k_{2} d k_{2} \,\overline{n}_{k_{2}}(\tau)j_{0}(k_{2} r)}{\int k_{1} d k_{1} \overline{n}_{k_{1}}(\tau)\, \int k_{2} d k_{2} \,\overline{n}_{k_{2}}(\tau)}.
\label{AMB5a}
\end{equation}
In the large-scale limit Eq. (\ref{AMB5a}) will then imply that $g^{(2)}(\vec{r},\tau) \to 2$ \cite{mandel}; conversely 
Eq. (\ref{CC11}), in the same physical limit, implies $g^{(2)}(\vec{r},\tau) \to 3$. 

\subsection{Why are we so close to Bose-Einstein correlations?}
The curvature quanta follow a Bose-Einstein multiplicity distribution\footnote{This occurrence can be traced back to a property of the $SU(1,1)$ group. The connection between the multiplicity distributions and the SU(1,1) group structure can be neatly expressed by computing, in explicit terms, the Wigner matrix element of the positive discrete series (see, in this respect, Ref. \cite{multmax})}. This is, in a nutshell, the reason why the degree of second order coherence of curvature inhomogeneities is very close to the Bose-Einstein case (i.e. $3$ rather than $2$).  
Consider the case of a two-mode squeezed state which corresponds to a single $\vec{k}$-mode of the field. We are interested in computing $\langle m\, m\, | \zeta\rangle$ where $|\zeta \rangle = \Sigma(\zeta) | 0\, 0\rangle$ 
and $\Sigma(\zeta)$ is actually given in Eq. (\ref{AA22}). We will have that the operator
$\Sigma(\zeta)$ can be factorized as \cite{schum}:
\begin{equation}
\Sigma(\zeta) = \exp{\biggl[ - \frac{\zeta}{|\zeta|} \tanh{|\zeta|} \, K_{+}\biggr]}\times \exp{[- 2 \ln{\cosh|\zeta|}\, K_{0}]}\times \exp{\biggl[ \frac{\zeta^{*}}{|\zeta|} \, \tanh{|\zeta|} K_{-}\biggr]},
\end{equation}
where $K_{+}$, $K_{-}$ and $K_{0}$ obey the standard commutation relations of the $SU(1,1)$ Lie algebra\footnote{Recalling the explicit expressions 
of Eq. (\ref{AA21}) it is simple to show that $[ K_{-},\, K_{+}] = 2\, K_{0}$ and that $[K_{0},\,K_{\pm}] = \pm K_{\pm}$.}. Using the previous equation we will have that
the multiplicity distribution is given by:
\begin{equation}
|\langle m\, m\,| \zeta \rangle|^2 = \frac{1}{\overline{n} + 1} \biggl(\frac{\overline{n}}{\overline{n} + 1} \biggr)^{m},\qquad \overline{n} = \sinh^2{r}.
\end{equation}
If we now take into account that we have a two-mode state {\em for each Fourier mode} the form of the 
density matrix becomes
\begin{equation}
\hat{\rho}_{\vec{k}} = \frac{1}{\cosh^2{r_{k}}} \sum_{n_{\vec{k}}=0}^{\infty} \sum_{m_{\vec{k}}=0}^{\infty} e^{- i \alpha_{k}(n_{\vec{k}} - m_{\vec{k}})} 
(\tanh{r_{k}})^{n_{\vec{k}} + m_{\vec{k}}} |n_{\vec{k}}\,\, n_{- \vec{k}}\rangle \langle m_{-\vec{k}}\,\,m_{\vec{k}}|,
\label{QM2a}
\end{equation}
whose diagonal elements define the multiplicity distribution which has precisely the Bose-Einstein form:
\begin{equation}
P_{\{n_{\vec{k}}\}} = \prod_{\vec{k}} P_{n_{\vec{k}}} , \qquad P_{n_{\vec{k}}}(\overline{n}_{k})= 
\frac{\overline{n}_{k}^{n_{\vec{k}}}}{( 1 + \overline{n}_{k})^{n_{\vec{k}} + 1 }},
\label{QM2}
\end{equation} 
accounting for the way curvature quanta of each Fourier mode (i.e. $n_{\vec{k}}$) 
are distributed as a function of their average multiplicity (i.e.  $\overline{n}_{k} = \sinh^2{r_{k}}$).  

The behaviour of the off-diagonal elements of Eq. (\ref{QM2a}) is dictated by the phases  
$\alpha_{k}= (2 \varphi_{k} - \gamma_{k})$ whose evolution is specifically discussed in appendix \ref{APPC} (see, in particular, Eq. (\ref{AA24}) and discussion therein). This explains the nature of the multiplicity distribution which is of 
Bose-Einstein type but with a larger amount of correlations (hence the value 
$3$ instead of $2$ for the normalized degree of second-order coherence). Such a difference also 
persists to higher order. By appropriately reducing the density 
matrix, the multiplicity distribution can become exactly  
to the one of a thermal state\footnote{Even if this point might be related to the dynamical evolution of $\alpha_{k}$ (see Eq. (\ref{AA24}) and discussion therein) we just point out that by averaging over $\alpha_{k}$ Eq. (\ref{QM2a}), 
 the density matrix can be reduced (i.e. $\hat{\rho}^{\mathrm{red}}_{\vec{k}} = \frac{1}{2\pi} \int_{0}^{2\pi} d \alpha_{k} \hat{\rho}_{\vec{k}}$) to a 
the density matrix of a chaotic state. A similar observation has been made long ago in a related context\cite{yurke}.}.
 
\subsection{The quantum mechanical correspondence}
The numerical values of the degrees of second-order coherence 
in the large-scale limit are solely determined by the statistical properties of the quantum state. This 
statement can be made more precise: the degrees of first- and second-order coherence 
in the large-scale limit coincide with the results obtainable if only a single mode of the field is excited in the limit of zero time delay 
between the quantum operators\footnote{In actual interferometry the electric field is first split into two components 
through the beam splitter, then it is time-delayed and finally recombined at the correlator. The limit of zero time delay between the signals is commonly used, in both cases,  to characterize the statistical properties of the source. We are here using a similar logic.}.  

Indeed, for a single mode of the field the degrees of first- and second-order coherence are defined as:
\begin{eqnarray}
\overline{g}^{(1)}(\tau_{1}, \tau_{2}) &=& \frac{\langle \hat{a}^{\dagger}(\tau_{1}) \, \hat{a}(\tau_{2})\rangle}{\sqrt{\langle \hat{a}^{\dagger}(\tau_{1}) \, \hat{a}(\tau_{1})\rangle}
\, \sqrt{\langle \hat{a}^{\dagger}(\tau_{2}) \, \hat{a}(\tau_{2})\rangle}},
\label{g1qm}\\
\overline{g}^{(2)}(\tau_{1}, \tau_{2}) &=& \frac{\langle \hat{a}^{\dagger}(\tau_{1})  \hat{a}^{\dagger}(\tau_{2}) \, \hat{a}(\tau_{2})\, \hat{a}(\tau_{1})\rangle}{\langle \hat{a}(\tau_{1}) \, \hat{a}(\tau_{1})\rangle \langle a^{\dagger}(\tau_{2}) \, a(\tau_{2})\rangle},
\label{g2qm}
\end{eqnarray}
where the overline at the left hand side distinguishes Eqs. (\ref{g1qm}) and (\ref{g2qm}) 
from Eqs. (\ref{corr6a}) and (\ref{g1})--(\ref{g2}) holding in the general case. 
Equations (\ref{g1qm}) and (\ref{g2qm}) define, respectively, the degrees of first and second-order temporal coherence: in the zero time-delay limit 
$\tau_{1}- \tau_{2}\to 0$ and in this case the degree of second-order coherence will be denoted by 
$\overline{g}^{(2)}$. Recalling the analysis of section \ref{sec4} we have already demonstrated, in practice, that:
\begin{equation}
\lim_{k r \ll 1} g^{(2)}(r, \tau) = \overline{g}^{(2)}.
\label{qm3}
\end{equation}
Equation (\ref{qm3}) is verified in the case of a single-mode coherent state (i.e. $\hat{a} |\alpha \rangle = \alpha  |\alpha \rangle$) where Eqs. (\ref{g1qm}) and (\ref{g2qm}) imply $\overline{g}^{(1)} = \overline{g}^{(2)} =1$.  Similarly for a chaotic state with statistical weights provided by the Bose-Einstein distribution the density matrix has the same form quoted before but for a single mode. Therefore we have that 
$\overline{g}^{(1)}=1$ but $\overline{g}^{(2)}=2$.  In the case of a single Fock state $\overline{g}^{(2)} = (1 - 1/n )< 1$ showing that Fock states lead always to sub-Poissonian behaviour and they are anti-bunched \cite{mandel,loudon}.  
Finally a single-mode of the field in the quasi-de Sitter case would correspond to $\hat{a} = \cosh{r} \hat{b} - \sinh{r} \hat{b}^{\dagger}$ where 
the phases have been fixed to zero; taking the limit of zero time-delay and inserting these expressions in Eq. (\ref{g2qm}) we have that: 
$\overline{g}^{(2)} = 3 + 1/\overline{n}$ with $\overline{n} = \sinh^2{r}$. In the quasi-de Sitter case the quantum correspondence 
holds provided the number of produced quanta in each Fourier mode is large so that terms of the order of $1/\overline{n}$ (and higher) 
are all negligible since $\overline{n}$. In this $1/\overline{n}$ expansion  the normal-ordered expectation values dominate the degrees of coherence  (see also the last part of appendix \ref{APPC}).
Chaotic light is an example of bunched 
 quantum state (i.e. $\overline{g}^{(2)}(0) > 1$ implying more degree of second-order coherence 
 than in the case of a coherent state). Fock states are instead antibunched (i.e. $\overline{g}^{(2)} <1$) implying 
 a degree of second-order coherence smaller than in the case of a coherent state. 
The correspondence of Eq. (\ref{qm3}) holds as well for the higher-order degrees of coherence, as we shall see. 

In the zero time-delay limit the degree of second-order coherence has a simple relation with the variance of the probability distribution
associated with a given quantum state.  More specifically Eq. (\ref{g2qm}) implies 
$ \overline{g}^{(2)} = (D^2 - \overline{n})/\overline{n}^2$ where $\overline{n} = \langle \hat{N}\rangle $ and $D^2 = \langle \hat{N}^2 \rangle 
- \langle \hat{N} \rangle^2$.  It is sometimes useful to define the so-called Mandel parameter whose expression is 
${\mathcal Q} = \overline{n} [ \overline{g}^{(2)} -1] = D^2/\overline{n} -1$.
The parameter ${\mathcal Q}$ is directly related to the way second-order correlations are parametrized in subatomic physics \cite{revs}. 
In the case of a coherent state \cite{glauber} we have that  $D^2 = \langle \hat{N} \rangle$ while ${\mathcal Q}=0$; this means that the distribution 
underlying this state is just the Poisson distribution (i.e. the variance coincides with the mean value). 
 
\subsection{Correlations from compound Poisson processes}
In the perspective of the present investigation the degrees of first and second-order coherence are not sufficient to disambiguate 
the statistical properties of the quantum state and need to be complemented by the analysis of some higher-order degrees 
of coherence.  We shall now show, in a specific class of examples, that very different quantum states lead to the same degrees of first-order and second-order coherence.  

Let us suppose that the quantum state of cosmological perturbations 
is unknown but characterized by a given density matrix whose statistical weights will be denoted by $P_{n}$, i.e. 
\begin{equation}
\hat{\rho} = \sum_{n=0}^{\infty} P_{n} \, | n\,\rangle \langle n |,\qquad \sum_{n =0}^{\infty} \, P_{n} =1.
\label{AMB0}
\end{equation}
We work in the zero-delay limit and exploit the results of Eqs. (\ref{g1qm}), (\ref{g2qm}) and (\ref{qm3}). 
The degrees of correlations can then be expressed as\footnote{Notice that Eqs. (\ref{AMB9})--(\ref{AMB13}) can be derived easily by recalling 
that, as usual, the expectation value of a give operator $\hat{O}$ is  given by 
$\langle \hat{O} \rangle = \mathrm{Tr}[ \hat{O} \, \hat{\rho}]$.}:
\begin{eqnarray}
\overline{g}^{(1)} &=& \frac{\langle \hat{a}^{\dagger} \, \hat{a} \rangle}{\overline{n}} = \frac{1}{\overline{n}} \frac{ d {\mathcal A}}{d s}\biggl|_{s=1} ,
\label{AMB9}\\
\overline{g}^{(2)} &=& \frac{\langle \hat{a}^{\dagger} \, \hat{a}^{\dagger} \hat{a}\, \hat{a} \rangle}{\overline{n}^2} = \frac{1}{\overline{n}^2} \frac{ d^2 {\mathcal A}}{d s^2}\biggl|_{s=1} , 
\label{AMB10}\\
\overline{g}^{(3)} &=& \frac{\langle \hat{a}^{\dagger}\, \hat{a}^{\dagger} \, \hat{a}^{\dagger} \hat{a}\, \hat{a} \, \hat{a} \rangle}{\overline{n}^3} = \frac{1}{\overline{n}^3} \frac{ d^3 {\mathcal A}}{d s^3}\biggl|_{s=1}, 
\label{AMB11}\\
\overline{g}^{(4)} &=& \frac{\langle \hat{a}^{\dagger}\, \hat{a}^{\dagger}\, \hat{a}^{\dagger} \, \hat{a}^{\dagger} \hat{a}\, \hat{a}\,  \hat{a}\, \hat{a}  \rangle}{\overline{n}^4} = \frac{1}{\overline{n}^4} \frac{ d^4 {\mathcal A}}{d s^4}\biggl|_{s=1}, 
\label{AMB12}
\end{eqnarray}
where $\overline{n} = \mathrm{Tr} [ \hat{a}^{\dagger} \, \hat{a} \, \hat{\rho}]$; the function ${\mathcal A}(s)$ appearing in Eqs. (\ref{AMB9})--(\ref{AMB12}) is the probability generating function defined as:
\begin{equation}
{\mathcal A}(s) = \sum_{n=0}^{\infty} \, s^{n} \, P_{n} , \qquad {\mathcal A}(1) = 1. 
\label{AMB13}
\end{equation}
One of the simplest situations coming to mind is the one where $P_{n}$ is a Poisson distribution. In this case we have that 
\begin{equation}
P_{n} = \frac{e^{- \overline{n}}}{n!} \, \overline{n}^{n}, \qquad {\mathcal A}(s) = e^{\overline{n}(s-1)}
\label{PP1}
\end{equation}
and thanks to Eqs. (\ref{AMB9})--(\ref{AMB13}) it is immediate to verify that the degree 
of nth-order coherence in the case of a Poisson distribution is always $1$. Note that Eq. (\ref{PP1}) 
does not define a coherent state but rather a mixed state with Poisson distribution. Another possible case which can be easily handled with Eqs. (\ref{AMB9})--(\ref{AMB13}) is when $P_{n}$ is a Bose-Einstein distribution. In this case $\overline{g}^{(n)} = n!$ and, as already discussed, 
$\overline{g}^{(2)} = 2$.

To reproduce the degree of second-order coherence of cosmological perturbations the simplest idea is to deviate 
slightly from the Bose-Einstein distribution and to consider the sum $A_{{\mathcal N}}$ of a number ${\mathcal N}$ of
mutually independent random variables, i.e.
\begin{equation}
A_{{\mathcal N}} = X_{1}+X_{2} +\,....+\, X_{\mathcal N}.
\label{TH1}
\end{equation}
If ${\mathcal N}$ is {\em fixed} the probability generating function ${\mathcal P}(s)$ of the sum $A_{{\mathcal N}}$ is simply given by the product of the generating functions. If ${\mathcal N}$ is itself a random number the generating function is given by the compound distribution.
Suppose that  $X_{i}$ are identically and independently 
distributed random variables with probability distribution $P_{k}$ and 
generating function ${\mathcal P}(s)$. If the distribution of ${\mathcal N}$ is 
$G_{k}$ its corresponding generating function will be, as usual 
${\mathcal G}(s) =  \sum_{k} s^{k} G_{k}$. The generating function of describing the compound process of Eq. (\ref{TH1})
will then be given by ${\mathcal A}(s) = {\mathcal G}[{\mathcal P}(s)]$ \cite{barucha}.

If ${\mathcal N}$ is distributed as a Poissonian variable with multiplicity 
$\overline{N}_{\mathrm{c}}$ the probability generating function will be given by $\exp{\{\overline{N}_{c}[{\mathcal P}(s) -1]\}}$.  If we take ${\mathcal P}(s)$ to be the generating function of a logarithmic distribution, i.e. 
\begin{equation}
{\mathcal P}(s)= 1 - \frac{k}{\overline{N}_{c}}\ln{\biggl[\frac{\overline{n}}{k}( 1- s) +1\biggr]},
\label{TH5}
\end{equation}
The generating function of the random sum of random variables will then be given by:
\begin{equation}
{\mathcal A}(s) = e^{\overline{N}_{c}[{\mathcal P}(s) -1]} = \frac{k^{k}}{[(1-s) \overline{n} + k]^{k}}, \qquad \overline{g}^{(n)}= \frac{k (k +1)\,.\,.\,.\,(k + n-1)}{k^{n}}.
\label{AMB14}
\end{equation}
In the case $k =1$ Eq. (\ref{AMB14}) reproduces the probability generating function of a Bose-Einstein distribution while, in the limit $k \to \infty$ 
${\mathcal A}(s)$ becomes exactly the Poisson generating function. Consider now, for the sake of concreteness, the case $k =1/2$:  the normalized degrees of first and second-order coherence reproduce the ones of a squeezed state to leading order in $1/\overline{n}$, namely from Eq. (\ref{AMB14})
\begin{equation}
\overline{g}^{(1)} =1, \qquad \overline{g}^{(2)} = 1 +\frac{1}{k} \to 3, \qquad k =1/2.
\label{AMB15c}
\end{equation}
Even assuming the the initial state of cosmological perturbations is exactly 
the state that minimizes the Hamiltonian of the fluctuations, the degree of second-order coherence 
can be easily  reproduced by the density matrix of an appropriately  mixed state whose statistical weights correspond 
to a compound Poisson process where each of the independent random variables of the sum follow a logarithmic 
distribution.  The probability generating function of Eq. (\ref{AMB14}) appears in the so-called 
pure death stochastic process when the evolution equation for the probability distributions are solved for 
a specific set of initial data \cite{barucha}. In this case, however, the interpretation 
of Eq. (\ref{AMB14}) would be slightly different and the value of $k$ bound to be integer.

\renewcommand{\theequation}{5.\arabic{equation}}
\setcounter{equation}{0}
\section{Higher-order correlations}
\label{sec5}
The degree of second-order coherence should  be complemented by some informations 
on the higher-order correlators. In this way the statistical properties of the quantum state 
can be disambiguated.

The importance  of the higher-order correlation functions of the detector's readings
has been emphasized in \cite{opt1} with the purpose of determining the coherence of the states of the field in an optical cavity.
Similarly in \cite{opt2} the third- and fourth-order autocorrelation functions have been used to detect the nonclassical character of the light transmitted through a photonic-crystal nanocavity containing a strongly coupled quantum dot probed with a train of coherent light pulses. The value of $\overline{g}^{(3)}$ has been 
contrasted with the conventional diagnostic based on $\overline{g}^{(2)}$. It has been demonstrated that, in addition to being necessary for detecting two-photon states emitted by a low-intensity source, $\overline{g}^{(3)}$ provides a more clear indication of the nonclassical character of a light source. In \cite{opt2} preliminary data have been presented to demonstrate bunching in the fourth-order autocorrelation function $\overline{g}^{(4)}$ as the first step toward detecting three-photon states.  Higher-order autocorrelations are necessary to characterize the multiphoton nature of nonclassical light also in the context of the so called quantum dots \cite{opt3,opt4}. Because of their strong interaction with light and ease of integration into optoelectronic devices, self-assembled quantum dots are promising candidates for quantum light sources, i.e. light sources leading, in our language, to $\overline{g}^{(2)} \ll 1$. All in all we can say that the specific analysis of $\overline{g}^{(2)}$ is often not sufficient to determine the statistical properties of light so that higher-order correlations have been recently analyzed \cite{opt3,opt4}.

The logic of the HBT interferometry will now be generalized by considering the correlations of three and four intensities in quasi-de Sitter space.
Recalling Eq. (\ref{corr5}) in the case $n=3$ the third order correlator corresponding to the three-point function of the 
intensity of the field evaluated at the same conformal time is given by:
\begin{eqnarray}
{\mathcal G}^{(3)}(\vec{x}_{1},\, \vec{x}_{2},\, \vec{x}_{3},\, \tau) &=&\frac{1}{8} \int \frac{d^{3} k_{1}}{(2\pi)^{3} \, k_{1}} .\,.\,.\, \int \frac{d^{3} k_{6}}{(2\pi)^{3} \, k_{6}}
\nonumber\\
&\times& e^{- i (\vec{k}_{1} + \vec{k}_{4}) \cdot \vec{x}_{1}} \, e^{- i (\vec{k}_{1} + \vec{k}_{3}) \cdot \vec{x}_{2}} \, e^{- i (\vec{k}_{3} + \vec{k}_{6}) \cdot \vec{x}_{3}}
\nonumber\\
&\times& \biggl[ \langle \hat{b}_{\vec{k}_{1}}\,  \hat{b}_{-\vec{k}_{2}} \hat{b}_{\vec{k}_{3}} \hat{b}_{-\vec{k}_{4}}^{\dagger}\,  \hat{b}_{-\vec{k}_{5}}^{\dagger} \hat{b}_{-\vec{k}_{6}}^{\dagger}\rangle \,v_{k_{1}}^{*} u_{k_{2}}^{*} v_{k_{3}}^{*} u_{k_{4}} v_{k_{5}} v_{k_{6}}
\nonumber\\
&+&  \langle \hat{b}_{\vec{k}_{1}}\,  \hat{b}_{-\vec{k}_{2}}^{\dagger} \hat{b}_{\vec{k}_{3}} \hat{b}_{\vec{k}_{4}}\,  \hat{b}_{-\vec{k}_{5}}^{\dagger} \hat{b}_{-\vec{k}_{6}}^{\dagger}\rangle
\,v_{k_{1}}^{*} v_{k_{2}}^{*} v_{k_{3}}^{*} v_{k_{4}} v_{k_{5}} v_{k_{6}}
\nonumber\\
&+& \langle \hat{b}_{\vec{k}_{1}}\,  \hat{b}_{\vec{k}_{2}} \hat{b}_{-\vec{k}_{3}}^{\dagger} \hat{b}_{\vec{k}_{4}}^{\dagger}\,  \hat{b}_{-\vec{k}_{5}}^{\dagger} \hat{b}_{-\vec{k}_{6}}^{\dagger}\rangle \, v_{k_{1}}^{*} v_{k_{2}}^{*} u_{k_{3}}^{*} u_{k_{4}} v_{k_{5}} v_{k_{6}}
\nonumber\\
&+& \langle \hat{b}_{\vec{k}_{1}}\,  \hat{b}_{-\vec{k}_{2}}^{\dagger} \hat{b}_{\vec{k}_{3}} \hat{b}_{-\vec{k}_{4}}^{\dagger}\,  \hat{b}_{\vec{k}_{5}} \hat{b}_{-\vec{k}_{6}}^{\dagger}\rangle\, v_{k_{1}}^{*} u_{k_{2}}^{*} v_{k_{3}}^{*} v_{k_{4}} u_{k_{5}} v_{k_{6}}\biggr],
\label{THIRD1}
\end{eqnarray}
where the six points appearing in Eq. (\ref{corr5}) in the case $n=3$ have been identified using the same procedure already 
explained when passing from Eq. (\ref{CC6}) to Eq. (\ref{CC6a}). The various correlators have been made explicit by eliminating three out the six integrals over the momenta thanks to the three delta functions arising from each of the four expectation values appearing at the right hand side of Eq. (\ref{THIRD1}). The result of this manipulation is given by:
\begin{eqnarray}
&&{\mathcal  G}^{(3)}(\vec{x}_{1},\, \vec{x}_{2},\, \vec{x}_{3},\, \tau)  =\frac{1}{8} \int \frac{d^{3} k_{1}}{(2\pi)^{6} \, k_{1}} \int \frac{d^{3} k_{2}}{(2\pi)^{6} \, k_{2}} 
\int \frac{d^{3} k_{3}}{(2\pi)^{6} \, k_{3}}  |v_{k_{1}}(\tau)|^2 |v_{k_{2}}(\tau)|^2 |v_{k_{3}}(\tau)|^2 
\nonumber\\
&&\times \biggl\{\biggl[ 1 + e^{- i (\vec{k_{2}}- \vec{k}_{3}) \cdot \vec{r}_{23} } + e^{- i (\vec{k}_{1} - \vec{k}_{2}) \cdot \vec{r}_{12}} + e^{- i (\vec{k}_{1} - \vec{k}_{3}) \cdot \vec{r}_{13}} + e^{ - i (\vec{k}_{1} \cdot\vec{r}_{12} + \vec{k}_{2}\cdot \vec{r}_{23} +\vec{k}_{3}\cdot \vec{r}_{31})} 
\nonumber\\
&&+ e^{ - i (\vec{k}_{1} \vec{r}_{13} + \vec{k}_{2} \vec{r}_{21} + \vec{k}_{3} \cdot \vec{r}_{32})}\biggr]
\nonumber\\
&&+ v_{k_{1}}^{*}(\tau) u_{k_{1}}^{*}(\tau) |v_{k_{3}}(\tau)|^2 u_{k_{2}}(\tau) v_{k_{2}}(\tau) \biggl[ e^{- i (\vec{k}_{1}\cdot \vec{r}_{12} + \vec{k}_{3} \cdot \vec{r}_{32} + \vec{k}_{2} \cdot \vec{r}_{13})} + e^{- i (\vec{k}_{1} + \vec{k}_{2})\cdot \vec{r_{12}}}\biggr]
\nonumber\\
&&+ v_{k_{2}}^{*}(\tau) u_{k_{2}}^{*}(\tau) |v_{k_{1}}(\tau)|^2 u_{k_{3}}(\tau) v_{k_{3}}(\tau)\biggr[ e^{ - i(\vec{k}_{1} \cdot \vec{r}_{12} + \vec{k}_{2} \cdot \vec{r}_{23} + \vec{k}_{3} \cdot \vec{r}_{13})} 
\nonumber\\
&& + e^{ - i[ ( \vec{k}_{1} + \vec{k}_{3}) \cdot \vec{x}_{1} - (\vec{k}_{2} + \vec{k}_{3})\cdot\vec{x}_{2}  - (\vec{k}_{1} + \vec{k}_{2})\cdot\vec{x}_{3}]}+ e^{- i (\vec{k}_{1} + \vec{k}_{2})\cdot\vec{r}_{13} }+ e^{- i (\vec{k}_{1}\cdot\vec{r}_{13}
+ \vec{k}_{2}\cdot\vec{r}_{23} + \vec{k}_{3}\cdot\vec{r}_{12})} 
\nonumber\\
&&+ e^{ - i [\vec{k}_{1}\cdot \vec{r}_{12} + \vec{k}_{2}\cdot \vec{r}_{31} - \vec{k}_{3}\cdot (\vec{x}_{2} + \vec{x}_3)]} \biggr]\biggr\},
\label{THIRD2}
\end{eqnarray}
where, by definition, $\vec{r}_{ij} = \vec{x}_{i} - \vec{x}_{j}$. 
The degree of third order coherence can be immediately computed and it is: 
\begin{equation}
g^{(3)}(\vec{x}_{1}, \vec{x}_{2}, \vec{x}_{3}, \tau) = \frac{{\mathcal G}^{(3)}(\vec{x}_{1},\, \vec{x}_{2},\, \vec{x}_{3},\, \tau) }{{\mathcal G}^{(1)}(\vec{x}_{1},\, \tau)\, 
{\mathcal G}^{(1)}(\vec{x}_{2},\, \tau)\, {\mathcal G}^{(1)}(\vec{x}_{3},\, \tau)}.
\label{THIRD2a}
\end{equation}
By evaluating Eq. (\ref{THIRD2a}) in the large-scale limit,  after some algebra the above result can be expressed as
\begin{equation}
\frac{\int d^{3}k_{1}/k_{1} \int d^{3}k_{2} /k_{2} \int d^{3}k_{3}/k_{3} |v_{k_{1}}(\tau)|^2 [ 5 |v_{k_{2}}(\tau)|^2 |v_{k_{3}}(\tau)|^2 + 6 v_{k_{2}}^{*}(\tau) u_{k_{2}}^{*}(\tau) 
v_{k_{3}}(\tau) u_{k_{3}}(\tau) ]}{\int d^{3} k_{1} |v_{k_{1}}(\tau)|^2/k_{1} \, \int d^{3} k_{2} |v_{k_{2}}(\tau)|^2/k_{2}\,\int d^{3} k_{3} |v_{k_{3}}(\tau)|^2/k_{3}}.
\label{THIRD3}
\end{equation}
It can be observed that the integral $\int d^{3} k_{1} |v_{k_{1}}(\tau)|^2/k_{1}$ appearing in the denominator of Eq. (\ref{THIRD3}) 
can be factorized in the numerator so that the expression for the degree of third-order coherence can be finally 
expressed as: 
\begin{equation}
\frac{ \int d^{3}k_{2} /k_{2} \int d^{3}k_{3}/k_{3} |v_{k_{1}}(\tau)|^2 [ 5 |v_{k_{2}}(\tau)|^2 |v_{k_{3}}(\tau)|^2 + 6 v_{k_{2}}^{*}(\tau) u_{k_{2}}^{*}(\tau) 
v_{k_{3}}(\tau) u_{k_{3}}(\tau) ]}{\int d^{3} k_{2} |v_{k_{2}}(\tau)|^2/k_{2}\,\int d^{3} k_{3} |v_{k_{3}}(\tau)|^2/k_{3}}
\label{THIRD3a}
\end{equation}
In the large-scale limit, the degree of third-order coherence becomes therefore:
\begin{equation}
\lim_{|k| r_{ij} \ll 1} g^{(3)}(\vec{x}_{1}, \vec{x}_{2}, \vec{x}_{3}) \to 11.
\label{THIRD3aa} 
\end{equation}
The reason why this is a numerical value and does not depend on the momenta shows that, in this limit, the degree of coherence 
only depends on the statistical properties of the quantum state, as explained in section \ref{sec4}. Indeed, Eq. (\ref{THIRD3a}) coincides with the result obtainable in the case of a single degree of freedom (recall the discussion after Eqs. (\ref{g1qm}) and (\ref{g2qm})) neglecting the higher orders in the $1/\overline{n}$ expansion:
\begin{equation}
 \overline{g}^{(3)} = \frac{\langle \hat{a}^{\dagger}\, \hat{a}^{\dagger} \, \hat{a}^{\dagger} \hat{a}\, \hat{a} \, \hat{a} \rangle}{\overline{n}^3}= 11 + \frac{4}{\overline{n}} + \frac{5}{\overline{n}^2}.
\label{AMB15}
\end{equation}
\begin{table}
\begin{center}
\vskip 0.5truecm
\begin{tabular}{| c | | |c | l | c | | | c | | |  c | | | c |  }
\hline
{\rm Process} & $k$ \quad & $\overline{g}^{(1)}$ & $\overline{g}^{(2)}$  &  $\overline{g}^{(3)}$& $\overline{g}^{(4)}$\\ \hline
{\rm Poisson Compound} & $1/2$ &  $1 (1)$ & $3 (3)$ & $15$ $(11)$& $105$ $(93)$ \\ \hline
{\rm Bose-Einstein} & $1$ &  $1$&  $2 $& $6$ & $24$ \\ \hline
{\rm Poisson} & $\infty$ & $1$ &  $1$  & $1$& $1$ \\ \hline
\hline
\end{tabular}
\caption{The degrees of coherence in various processes. In the first row we report inside the brackets the degrees of coherence computed for the quantum state of curvature perturbations.}
\label{Table1}
\end{center}
\end{table}

Finally going back to Eq. (\ref{corr5}) in the $n=4$ case, the fourth-order 
correlator is given by:
\begin{eqnarray}
{\mathcal G}^{(4)}(\vec{x}_{1},\, \vec{x}_{2},\, \vec{x}_{3},\, \vec{x}_{4},\, \tau) &=&\frac{1}{16} \int \frac{d^{3} k_{1}}{(2\pi)^{3} \, k_{1}} .\,.\,.\, \int \frac{d^{3} k_{8}}{(2\pi)^{3} \, k_{8}}
\nonumber\\
&\times& e^{- i (\vec{k}_{1} + \vec{k}_{3}) \cdot \vec{x}_{1}} \, e^{- i (\vec{k}_{2} + \vec{k}_{6}) \cdot \vec{x}_{2}} \, e^{- i (\vec{k}_{3} + \vec{k}_{7}) \cdot \vec{x}_{3}}\, e^{- i (\vec{k}_{4} + \vec{k}_{8}) \cdot \vec{x}_{4}}
\nonumber\\
&\times& \biggl[ \langle \hat{b}_{\vec{k}_{1}}\,  \hat{b}_{\vec{k}_{2}} \hat{b}_{\vec{k}_{3}} \hat{b}_{\vec{k}_{4}}  \hat{b}_{-\vec{k}_{5}}^{\dagger} \hat{b}_{-\vec{k}_{6}}^{\dagger}\hat{b}_{-\vec{k}_{7}}^{\dagger} \hat{b}_{-\vec{k}_{8}}^{\dagger}\rangle v_{k_{1}}^{*} v_{k_{2}}^{*} v_{k_{3}}^{*} v_{k_{4}}^{*} v_{k_{5}} v_{k_{6}} v_{k_{7}} v_{k_{8}}
\nonumber\\
&+&  \langle \hat{b}_{\vec{k}_{1}}\,  \hat{b}_{-\vec{k}_{2}}^{\dagger} \hat{b}_{\vec{k}_{3}} \hat{b}_{\vec{k}_{4}}  \hat{b}_{\vec{k}_{5}} \hat{b}_{-\vec{k}_{6}}^{\dagger}\hat{b}_{-\vec{k}_{7}}^{\dagger} \hat{b}_{-\vec{k}_{8}}^{\dagger}\rangle v_{k_{1}}^{*} u_{k_{2}}^{*} v_{k_{3}}^{*} v_{k_{4}}^{*} u_{k_{5}} v_{k_{6}} v_{k_{7}} v_{k_{8}}
\nonumber\\
&+& \langle \hat{b}_{\vec{k}_{1}}\,  \hat{b}_{\vec{k}_{2}} \hat{b}_{-\vec{k}_{3}}^{\dagger} \hat{b}_{\vec{k}_{4}} \hat{b}_{\vec{k}_{5}} \hat{b}_{-\vec{k}_{6}}^{\dagger}\hat{b}_{-\vec{k}_{7}}^{\dagger} \hat{b}_{-\vec{k}_{8}}^{\dagger}\rangle v_{k_{1}}^{*} v_{k_{2}}^{*} u_{k_{3}}^{*} v_{k_{4}}^{*} u_{k_{5}} v_{k_{6}} v_{k_{7}} v_{k_{8}}
\nonumber\\
&+& \langle \hat{b}_{\vec{k}_{1}}\,  \hat{b}_{\vec{k}_{2}} \hat{b}_{-\vec{k}_{3}}^{\dagger} \hat{b}_{\vec{k}_{4}}  \hat{b}_{-\vec{k}_{5}}^{\dagger} \hat{b}_{\vec{k}_{6}}\hat{b}_{-\vec{k}_{7}}^{\dagger} \hat{b}_{-\vec{k}_{8}}^{\dagger}\rangle v_{k_{1}}^{*} v_{k_{2}}^{*} v_{k_{3}}^{*} u_{k_{4}}^{*} u_{k_{5}} v_{k_{6}} v_{k_{7}} v_{k_{8}}
\nonumber\\
&+& \langle \hat{b}_{\vec{k}_{1}}\,  \hat{b}_{\vec{k}_{2}} \hat{b}_{\vec{k}_{3}} \hat{b}_{-\vec{k}_{4}}^{\dagger}  \hat{b}_{\vec{k}_{5}} \hat{b}^{\dagger}_{-\vec{k}_{6}}\hat{b}_{-\vec{k}_{7}}^{\dagger} \hat{b}_{-\vec{k}_{8}}^{\dagger}\rangle v_{k_{1}}^{*} v_{k_{2}}^{*} v_{k_{3}}^{*} u_{k_{4}}^{*} u_{k_{5}} v_{k_{6}} v_{k_{7}} v_{k_{8}}
\nonumber\\
&+& \langle \hat{b}_{\vec{k}_{1}}\,  \hat{b}_{\vec{k}_{2}} \hat{b}_{-\vec{k}_{3}}^{\dagger} \hat{b}_{-\vec{k}_{4}}^{\dagger}  \hat{b}_{\vec{k}_{5}} \hat{b}_{\vec{k}_{6}}\hat{b}_{-\vec{k}_{7}}^{\dagger} \hat{b}_{-\vec{k}_{8}}^{\dagger}\rangle v_{k_{1}}^{*} v_{k_{2}}^{*} u_{k_{3}}^{*} u_{k_{4}}^{*} u_{k_{5}} u_{k_{6}} v_{k_{7}} v_{k_{8}}
\nonumber\\
&+& \langle \hat{b}_{\vec{k}_{1}}\,  \hat{b}_{\vec{k}_{2}} \hat{b}_{\vec{k}_{3}} \hat{b}_{-\vec{k}_{4}}^{\dagger}  \hat{b}_{-\vec{k}_{5}}^{\dagger} \hat{b}_{\vec{k}_{6}}\hat{b}_{-\vec{k}_{7}}^{\dagger} \hat{b}_{-\vec{k}_{8}}^{\dagger}\rangle u_{k_{1}}^{*} u_{k_{2}}^{*} v_{k_{3}}^{*} u_{k_{4}}^{*} v_{k_{5}} u_{k_{6}} v_{k_{7}} v_{k_{8}}
\nonumber\\
&+& \langle \hat{b}_{\vec{k}_{1}}\,  \hat{b}_{\vec{k}_{2}} \hat{b}_{-\vec{k}_{3}}^{\dagger} \hat{b}_{-\vec{k}_{4}}^{\dagger}  \hat{b}_{\vec{k}_{5}} \hat{b}_{-\vec{k}_{6}}^{\dagger}\hat{b}_{\vec{k}_{7}} \hat{b}_{-\vec{k}_{8}}^{\dagger}\rangle v_{k_{1}}^{*} u_{k_{2}}^{*} v_{k_{3}}^{*} v_{k_{4}}^{*} u_{k_{5}} v_{k_{6}} u_{k_{7}} v_{k_{8}}
\nonumber\\
&+& \langle \hat{b}_{\vec{k}_{1}}\,  \hat{b}_{\vec{k}_{2}} \hat{b}_{\vec{k}_{3}} \hat{b}_{-\vec{k}_{4}}^{\dagger}  \hat{b}_{-\vec{k}_{5}}^{\dagger} \hat{b}_{-\vec{k}_{6}}^{\dagger}\hat{b}_{\vec{k}_{7}}\hat{b}_{-\vec{k}_{8}}^{\dagger}\rangle v_{k_{1}}^{*} v_{k_{2}}^{*} v_{k_{3}}^{*} u_{k_{4}}^{*} v_{k_{5}} v_{k_{6}} u_{k_{7}} v_{k_{8}}
\nonumber\\
&+& \langle \hat{b}_{\vec{k}_{1}}\,  \hat{b}_{-\vec{k}_{2}}^{\dagger} \hat{b}_{\vec{k}_{3}} \hat{b}_{-\vec{k}_{4}}^{\dagger}  \hat{b}_{\vec{k}_{5}} \hat{b}_{-\vec{k}_{6}}^{\dagger}\hat{b}_{\vec{k}_{7}}\hat{b}_{-\vec{k}_{8}}^{\dagger}\rangle v_{k_{1}}^{*} u_{k_{2}}^{*} v_{k_{3}}^{*} u_{k_{4}}^{*} u_{k_{5}} v_{k_{6}} u_{k_{7}} v_{k_{8}}\biggr]
\label{FOURTH1}
\end{eqnarray}
Equation (\ref{FOURTH1}) must be evaluated in the large-scale limit and the degree of fourth-order coherence becomes:
\begin{equation}
g^{(4)}(\vec{x}_{1}, \vec{x}_{2}, \vec{x}_{3},\vec{x}_{4} \tau) = \frac{{\mathcal G}^{(3)}(\vec{x}_{1},\, \vec{x}_{2},\, \vec{x}_{3},\, \tau) }{{\mathcal G}^{(1)}(\vec{x}_{1},\, \tau)\, 
{\mathcal G}^{(1)}(\vec{x}_{2},\, \tau)\, {\mathcal G}^{(1)}(\vec{x}_{3},\, \tau)}.
\label{FOURTH3}
\end{equation}
After making explicit all the expectation values and after rearranging the integrations over the momenta one of the integrals simplifies between the numerator and the denominator. Inserting the explicit expressions for $u_{k}(\tau)$ and $v_{k}(\tau)$ in the large-scale limit we have that 
\begin{equation}
\lim_{|k| r_{ij} \ll 1} g^{(4)}(\vec{x}_{1}, \vec{x}_{2}, \vec{x}_{3}) = \frac{{\mathcal N}^{(4)}(\tau)}{{\mathcal D}^{(4)}(\tau)} \to 93.
\label{FOURTH3aa} 
\end{equation}
where ${\mathcal N}^{(4)}(\tau)$ and ${\mathcal D}^{(4)}(\tau)$ are defined as
\begin{eqnarray}
{\mathcal N}^{(4)}(\tau) &=&\int\frac{d^{3}k_{1}}{k_{1}} \int\frac{d^{3}k_{2}}{k_{2}} \int\frac{d^{3}k_{3}}{k_{3}} \biggl[ 24 |v_{k_{1}}(\tau)|^2 |v_{k_{2}}(\tau)|^2 |v_{k_{3}}(\tau)|^2
\nonumber\\
&+& 62 \,v_{k_{1}}^{*}(\tau)  u^{*}_{k_{1}}(\tau) v_{k_{2}}(\tau) u_{k_{2}}(\tau) |v_{k_{3}}(\tau)|^2 +  7 |u_{k_{1}}(\tau)|^2 v_{k_{2}}^{*}(\tau) u_{k_{2}}^{*}(\tau) v_{k_{3}}(\tau) u_{k_{3}}(\tau) \biggr],
\nonumber\\
{\mathcal D}^{(4)}(\tau) &=&\int\frac{d^{3}k_{1}}{k_{1}}  |v_{k_{1}}(\tau)|^2  \int\frac{d^{3}k_{2}}{k_{2}}  |v_{k_{2}}(\tau)|^2 \int\frac{d^{3}k_{3}}{k_{3}}  |v_{k_{3}}(\tau)|^2. 
\label{FOURTH2}
\end{eqnarray}
Again Eq. (\ref{FOURTH3aa}) agrees with the quantum mechanical correspondence in the $1/\overline{n}$ expansion, i.e. $g^{(4)} \to \overline{g}^{(4)}$ where 
\begin{equation}
\overline{g}^{(4)} = \frac{\langle \hat{a}^{\dagger} \, \hat{a}^{\dagger}\, \hat{a}^{\dagger} \, \hat{a}^{\dagger} \hat{a}\, \hat{a} \, \hat{a}\, \hat{a} \rangle}{\overline{n}^4}= 93 + \frac{76}{\overline{n}} + \frac{7}{\overline{n}^2}.
\label{AMB16}
\end{equation}
The third and fourth-order degrees of coherence complement the informations coming from the standard HBT interferometry. 
This analysis shows that the statistical properties of the quantum states can then be disambiguated 
by examining the higher degrees of coherence as exemplified in Tab. \ref{Table1}. In the case of the compound Poisson process of section \ref{sec4} the degrees 
of first-order and second-order coherence coincide with the ones of large-scale curvature modes (see the case $k=1/2$ in Tab. \ref{Table1}). By going to higher order the results differ substantially (see the values in brackets in the last two column). In the second and third lines we illustrated, for comparison, the Bose-Einsteins case (i.e. $k \to 1$ in Eq. (\ref{AMB14})) and the Poisson case (i.e. $k \to \infty$ in Eq. (\ref{AMB14})).

\renewcommand{\theequation}{6.\arabic{equation}}
\setcounter{equation}{0}
\section{Observational perspectives}
\label{sec6}
 \subsection{Counting detectors for single CMB photons}
The considerations developed here suggest the interesting possibility that counting 
detectors for single CMB photons could provide the observational counterpart for the considerations developed in the 
previous sections. While it would naively seem totally hopeless to conceive and 
build this kind of detectors,  a variety of cryogenic  technologies are being developed \cite{nano,matsuo2,matsuo3} for CMB observation at THz ($1\, \mathrm{THz} = 10^{3} \mathrm{GHz}$) frequencies. Furthermore the idea suggested in Ref. \cite{max1} of applying the HBT interferometry to CMB photons has been taken seriously in \cite{matsuo2,matsuo3}. According to these 
references it is not unconceivable that, in the future,  bunched CMB photons could be identified and collected in the THz range. 

It is not our purpose to review here the ideas related to photon counting THz interferometry (PCTI in the language of Ref. \cite{matsuo3}) which is, in oversimplified terms, an advanced version of the HBT  interferometry \cite{matsuo3}. In what follows we shall instead focus the discussion 
on the definition of the various observables (i.e. the degrees of coherence). The sole purpose of these considerations will be to relate 
the degrees of quantum coherence to actual CMB observables. If we consider the curvature 
inhomogeneities as quantum fields, the brightness perturbations will also become operators.  More specifically the curvature perturbation ${\mathcal R}$ is related to the field operator $\hat{q}$ (see appendix \ref{APPA}). The intensity fluctuations of the radiation field (i.e. the warmer and the cooler regions in the CMB sky)  are given, in turn, by  $\Delta_{\mathrm{I}}(\vec{x},\tau_{\mathrm{dec}}) \simeq - {\mathcal R}(\vec{n},\tau_{\mathrm{dec}})/5$ under the approximation of sudden decoupling  (i.e. assuming that the visibility function is a narrow Gaussian centered at $\tau_{\mathrm{dec}})$ \cite{v1,v2}.  
This correspondence will now be swiftly outlined.

\subsection{Brightness perturbations}

Indeed the degrees of coherence can be directly assessed by studying higher order temperature and polarization correlations in the limit of large angular scales. In the concordance scenario with no tensors the temperature anisotropies are 
customarily described, in real space by $\Delta_{I}(\hat{n},\tau)$ (the brightness perturbation of the intensity). The polarization 
anisotropies  are instead discussed in terms of the $E$ and $B$-modes, denoted, respectively, by  $\Delta_{E}(\hat{n},\tau)$ and $\Delta_{B}(\hat{n},\tau)$. Both $\Delta_{E}$ and $\Delta_{B}$ are spin $0$ 
quantities, exactly as $\Delta_{I}$ \cite{sud}. Recalling that $\mu = \cos{\vartheta}$ we have, in the concordance framework and in the 
absence of gravitons,
\begin{equation}
\Delta_{\mathrm{E}}(\hat{n}, \tau) = -  \partial_{\mu}^{2} \{( 1 - \mu^2) 
\Delta_{\mathrm{P}}(\hat{n},\tau)\},\qquad \Delta_{\mathrm{B}}(\hat{n}, \tau) = 0.
\label{TP5}
\end{equation}
The brightness perturbations $\Delta_{I}(\hat{n},\tau)$ and $\Delta_{P}(\hat{n},\tau)$ obey, in Fourier space, the following pair of equations: 
\begin{eqnarray}
&& \Delta_{I}^{\prime} + (i k \mu + \varepsilon') \Delta_{I} = \psi^{\prime} - i k \mu \phi
 +  \varepsilon' \biggl[\Delta_{\mathrm{I}0} + \mu v_{\mathrm{b}} + \frac{(1 - 3 \mu^2)}{4}S_{\mathrm{P}}(k,\tau)\biggl],
\label{TP9}\\
&& \Delta_{P}^{\prime} + (i k \mu + \varepsilon') \Delta_{P} = \frac{3 \varepsilon^{\prime}}{4}(1 - \mu^2) S_{\mathrm{P}}(k,\tau),
\label{TP10}
\end{eqnarray}
where $S_{\mathrm{P}}(k,\tau)$ can be expressed as the sum of the 
quadrupole of the intensity, of the monopole of the polarization and 
of the quadrupole of the  polarization, i.e.,   respectively, $S_{\mathrm{P}}(k,\tau) = 
(\Delta_{\mathrm{I}2} + \Delta_{\mathrm{P}0} + \Delta_{\mathrm{P}2})$; note that, in Eqs. (\ref{TP9}) and (\ref{TP10}), 
$\varepsilon'$ and $\varepsilon(\tau,\tau_{0})$ denote, respectively,  the differential optical depth and the optical depth 
itself \cite{v1,v2}
\begin{equation}
\varepsilon'= x_{\mathrm{e}} \tilde{n}_{\mathrm{e}} \,a\,\sigma_{\gamma\mathrm{e}}, \qquad 
\varepsilon(\tau,\tau_{0}) = \int_{\tau_{0}}^{\tau}  x_{\mathrm{e}} \tilde{n}_{\mathrm{e}} \,a\,\sigma_{\gamma\mathrm{e}} d\tau.
\label{TP10a}
\end{equation}
The line of sight solution of Eqs. (\ref{TP9}) and (\ref{TP10}) can be written, respectively, as: 
\begin{eqnarray}
\Delta_{\mathrm{I}}(k, \mu, \tau_{0}) &=& \int_{0}^{\tau_{0}} {\mathcal K}(\tau) \biggl[ \Delta_{\mathrm{I}0} + \phi + \mu v_{\mathrm{b}}  + \frac{(1 - 3 \mu^2)}{4} S_{\mathrm{P}}\biggr] e^{- i \mu x(\tau)}
\nonumber\\
&+& \int_{0}^{\tau_{0}} d\tau e^{- \varepsilon(\tau,\tau_{0})} ( \phi' + \psi') e^{-i \mu x(\tau)} d\tau,
\label{TP11}\\
\Delta_{\mathrm{P}}(k,\mu,\tau_{0}) &=& \frac{3}{4}(1-\mu^2) \int_{0}^{\tau_{0}}{\mathcal K}(\tau) S_{\mathrm{P}}(k,\tau) e^{- i k\mu (\tau- \tau_{0})} d\tau,
\label{TP12}
\end{eqnarray}
where ${\mathcal K}(\tau) = \varepsilon' e^{- \varepsilon(\tau,\tau_{0})}$ is the visibility function 
and $x(\tau) = k (\tau_{0} - \tau)$. The visibility function can be approximated as 
a double Gaussian with two peaks roughly corresponding to the redshifts 
of recombination and reionization \cite{v1,v2}. In the present analysis the focus will be on the large-angular scales 
corresponding to  typical multipoles $\ell \leq \sqrt{z_{\mathrm{rec}}}$ where the finite width 
of the visibility function is immaterial and the opacity suddenly drops at recombination. This 
implies that the visibility function presents a sharp (i.e. infinitely thin peak at the recombination time).  Thus, since ${\mathcal K}(\tau)$ is proportional to a Dirac delta 
function and $e^{- \varepsilon(\tau,\tau_{0})}$ is proportional to an Heaviside theta function.  Under the latter approximations, 
 Eq. (\ref{TP9}) leads to the well known pair of separated contributions, i.e. the Sachs-Wolfe (SW) and the integrated Sachs-Wolfe  (ISW) contributions:
\begin{eqnarray}
 \Delta_{\mathrm{I}}(k,\mu,\tau_{0}) &=& \Delta_{\mathrm{I}}^{(\mathrm{SW})}(k,\mu,\tau_{0})  + \Delta_{\mathrm{I}}^{(\mathrm{ISW})}(k,\mu,\tau_{0}), 
\label{HT14a}\\
\Delta_{\mathrm{I}}^{(\mathrm{SW})}(k,\mu,\tau_{0}) &=& \biggl( - \frac{{\mathcal R}(k,\tau)}{5}\biggr)_{\tau_{\mathrm{rec}}} e^{- i \mu y_{\mathrm{rec}}},
\nonumber\\
\Delta_{\mathrm{I}}^{(\mathrm{ISW})}(k,\mu,\tau_{0}) &=& \int_{\tau_{\mathrm{rec}}}^{\tau_{0}} 
(\phi' +\psi') e^{- i \mu x(\tau)} \, d\tau,
\label{HT15}
\end{eqnarray}
where, by definition, $x(\tau_{\mathrm{rec}}) = y_{\mathrm{rec}}$.
  The SW and the ISW contributions of Eqs. (\ref{HT14a}) and (\ref{HT15}) can be separately evaluated. In particular, as anticipated,  the ordinary SW contribution becomes: 
\begin{eqnarray}
\Delta_{\mathrm{I}}^{(\mathrm{SW})}(k,\mu,\tau_{0}) &=& - \frac{{\mathcal R}(\vec{k},\tau_{\mathrm{i}})}{5} 
{\mathcal S}(q) e^{- i \mu y_{\mathrm{rec}}} ,
\label{TP20A}\\
{\mathcal S}(q) &=& 1 + \frac{4}{3 q} - \frac{16}{3 q^2}
+ \frac{16( \sqrt{q+1} -1)}{3 q^3},
\label{TP20B}
\end{eqnarray}
while the ISW contribution is:
\begin{eqnarray}
&&\Delta_{\mathrm{I}}^{(\mathrm{ISW})}(k,\mu,\tau_{0}) = - 2 {\mathcal R}(\vec{k},\tau_{\mathrm{i}}) \int_{\tau_{\mathrm{rec}}}^{\tau_{0}} {\mathcal T}^{\,\,\prime}_{\mathcal R}(\tau)e^{- i \mu x(\tau)} \, d\tau,
\label{TP20C}\\
&& {\mathcal T}_{\mathcal R}(\tau) = 1 - \frac{{\mathcal H}(\tau)}{a^2(\tau)} \int_{0}^{\tau} a^2(\tau')\, d\tau'.
\label{TP20D}
\end{eqnarray}
Both in Eqs. (\ref{TP20A}) and (\ref{TP20C}), ${\mathcal R}(\vec{k},\tau)$ denotes the constant value of curvature perturbations 
at $\tau_{\mathrm{i}} < \tau_{\mathrm{eq}}$. By further approximating the integrand in Eq. (\ref{TP20C}) the whole large-scale 
contribution can be written, for the present purposes, as 
\begin{equation}
\Delta_{\mathrm{I}}(k,\tau_{0}) =  - {\mathcal R}(\vec{k},\tau_{\mathrm{i}}) \biggl\{  \frac{{\mathcal S}(q)}{5} +  \biggl[{\mathcal T}^{\,\,\prime}_{\mathcal R} \biggr] \biggr\}_{q_{\mathrm{rec}}}\,e^{ - i \mu y_{\mathrm{rec}}},
\label{TP21}
\end{equation}
where $q_{\mathrm{rec}} = a_{\mathrm{rec}}/a_{\mathrm{eq}}  \simeq 3.04$ for $h_{0}^2 \Omega_{\mathrm{M}0} \simeq 0.134$ (as usual $\Omega_{\mathrm{M}0}$ denotes the total matter density in critical units at the present time). Note that if we neglect the ISW term and assume 
that $q\gg 1$ we obtain the conventional simplified form of the SW contribution.

If the curvature perturbations are described by a quantum field operator, also the brightness perturbations 
must be treated quantum mechanically. The degrees of coherence 
defined in Eqs. (\ref{g2}), (\ref{g3}) and (\ref{g4}) can therefore be written as:
\begin{eqnarray}
g^{(2)}(\hat{m},\hat{n}) &=& 
\frac{\langle \hat{\Delta}^{(-)}(\hat{m}) \, \hat{\Delta}^{(-)}(\hat{n})\,
\Delta^{(+)}(\hat{m}) \, \Delta^{(+)}(\hat{n})\rangle}{\langle  \hat{\Delta}^{(-)}(\hat{m}) \, \hat{\Delta}^{(+)}(\hat{m})\rangle \langle  \hat{\Delta}^{(-)}(\hat{n}) \, \hat{\Delta}^{(+)}(\hat{n}) \rangle},
\label{g2a}\\
g^{(3)}(\hat{m}, \hat{n}, \hat{r}) &=& \frac{\langle \hat{\Delta}^{(-)}(\hat{m}) \, \hat{\Delta}^{(-)}(\hat{n})\,\hat{\Delta}^{(-)}(\hat{r})
\Delta^{(+)}(\hat{m}) \, \Delta^{(+)}(\hat{n})\rangle}{\langle  \hat{\Delta}^{(-)}(\hat{m}) \, \hat{\Delta}^{(+)}(\hat{m})\rangle \langle  \hat{\Delta}^{(-)}(\hat{n}) \, \hat{\Delta}^{(+)}(\hat{n}) \rangle \langle \hat{\Delta}^{(-)}(\hat{n}) \hat{\Delta}^{(+)}(\hat{n})\rangle},
\label{g3a}
\end{eqnarray}
and similarly for $g^{(4)}(\hat{m}, \hat{n}, \hat{r}, \hat{s})$. In Eqs. (\ref{g2a}) and (\ref{g3a}) we decomposed the 
positive and negative frequency parts of the operators as $\hat{\Delta}(\hat{n})= \hat{\Delta}^{(+)}(\hat{n}) + \hat{\Delta}^{(-)}(\hat{n})$; note finally 
that $\hat{\Delta}(\hat{n})$ can denote, indifferently, $\hat{\Delta}_{I}$ of $\hat{\Delta}_{E}$ (recall that we are here discussing the case 
where the B-mode polarization is absent). This means that we can define separate degrees of coherence 
for the temperature, for the $E$-mode polarization and for their cross-correlation.

To give an example we can compute, for instance, the degree of second-order coherence of the temperature. By substituting $\hat{\Delta}$ with 
$\hat{\Delta}_{I}$ in Eq. (\ref{g2a}) we shall have that 
\begin{eqnarray}
g^{(2)}(\hat{m},\hat{n}) &=& 1 + \frac{\sum_{\ell,\,\ell^{\prime}} \int q |v_{q}|^2\,d q \int p |v_{p}|^2 \,d p  A_{\ell}(q,\hat{m}\cdot\hat{n})A_{\ell^{\prime}}(p,\hat{m}\cdot\hat{n} )}{\sum_{\ell}\int q d q |v_{q}|^2 A_{\ell}(q, \hat{m}\cdot\hat{n}) \,\, \sum_{\ell^{\prime}} \int p d p |v_{p}|^2  A_{\ell^{\prime}}(q, \hat{m}\cdot\hat{n})}
\nonumber\\
&+& \frac{\sum_{\ell,\,\ell^{\prime}} \int q v_{q}^{*}\,u_{q}^{*}d q \int p v_{p}\, u_{p} \,d p  A_{\ell}(q,\hat{m}\cdot\hat{n})A_{\ell^{\prime}}(p,\hat{m}\cdot\hat{n})}{\sum_{\ell}\int q d q |v_{q}|^2 A_{\ell}(q, \hat{m}\cdot\hat{n}) \,\, \sum_{\ell^{\prime}} \int p d p |v_{p}|^2  A_{\ell}(q, \hat{m}\cdot\hat{n})},
\nonumber\\
A_{\ell}(q,\hat{m}\cdot\hat{n}) &=& ( 2 \ell+1) j_{\ell}^2(k\tau_{0}) P_{\ell}(\hat{m}\cdot\hat{n}),
 \label{g2b}
\end{eqnarray}
where $j_{\ell}(k\tau_{0})$ are the spherical Bessel functions and $P_{\ell}(\hat{m}\cdot\hat{n})$ the Legendre polynomials.
Equation (\ref{g2b}) directly relates the curvature perturbations to the brightness perturbations. 
All the inequalities established for the degree of second-order coherence  and all the 
considerations presented before are also applicable to Eq. (\ref{g2b}): in the limit 
$\hat{m}\cdot\hat{n} \to 1$, using  well known identities, the degree of coherence 
 coincides with the results obtained before. We shall skip these discussions which can be 
 found elsewhere \cite{max1}.
 
\subsection{Multiplicity distributions and all-order observables} 
We conclude with some speculation concerning the multiplicity distributions of the {\em total} number of phonons. 
If observed these distributions might usefully complement the interferometric approach discussed 
in the previous subsection.

In the hypothesis that curvature quanta 
are the sole source of temperature inhomogeneities the distribution $P(n)$ of the {\em total} number of phonons 
$n = \sum_{\vec{k}} n_{\vec{k}}$ must reflect the distribution of the warmer and cooler regions. The multiplicity distribution $P(n)$ accounts for the way the total number 
of phonons $n$ is distributed as a function of its mean value; $P(n)$ 
can be very different from the $P_{\{n_{k}\}}$ which has been discussed in Eq. (\ref{QM2}).  
Denoting with $p(\{n\})$ the joint probability distribution of the set of phonon occupation numbers $\{n\}$ of the field, we shall have that \cite{mandel}
\begin{equation}
p(\{n\}) = \prod_{\vec{k}}\frac{1}{(1 + \overline{n}_{\vec{k}}) ( 1 + 1/\overline{n}_{\vec{k}})^{n_{\vec{k}}}}.
\label{QM4}
\end{equation}
For any mode for which 
$\overline{n}_{k} =0$, the corresponding factor must be interpreted  as $\delta_{n_{\vec{k}} \, 0}$. In the following 
we shall suppose, quite generally, that only a subset consisting of 
$\epsilon$ modes of the field is actually occupied and we shall restrict the attention to this subset of modes. 
If $n$ is the total number of phonons and $P(n)$ is the multiplicity distribution of $n$, then 
$P(m) = \sum_{\{n\}} p(\{n\}) \delta_{m \, n}$. In quantum optics an analog of the multiplicity distribution $P(m)$ 
describes the statistical properties of (unpolarized) chaotic light beams \cite{mandel}.  
The evaluation of $P(m)$ can be in general difficult but 
it becomes easy in the physical case when the average occupation number $\overline{n}_{\vec{k}}$ of all the $\epsilon$ occupied modes become equal\footnote{In the case of a thermal light beam  which is either fully polarized or fully unpolarized the use 
of a rectangular spectral density is an excellent approximation in the derivation of the 
photocounting statistics which is also experimentally accessible \cite{mandel}.}; in this case  the joint probability distribution of the occupied modes becomes:
\begin{equation}
p(\{n\}) = \frac{1}{(1 +  \overline{n}/\epsilon)^{\epsilon} ( 1 + \epsilon/\overline{n})^{n}}, \qquad \overline{n} = \sum_{\vec{k}} \overline{n}_{k} = \epsilon \,\overline{n}_{k}.
\label{QM7}
\end{equation}
Every non-vanishing term in the summation $P(m) = \sum_{\{n\}} p(\{n\}) \delta_{m \, n}$ has  the same 
value and the required probability 
$P(m)$ is simply $p(\{n\})$  given by Eq. (\ref{QM7}) multiplied by a well known combinatorial factor accounting 
for the way the $n$ phonons are distributed among the $\epsilon$ modes:
\begin{equation}
P_{n}(\overline{n},\epsilon)=  \frac{\Gamma( n + \epsilon)}{\Gamma(\epsilon) \Gamma(n + 1)} \biggl(\frac{\overline{n}}{\overline{n} + \epsilon}\biggr)^{n} \biggl(\frac{\epsilon}{\overline{n} + \epsilon}\biggr)^{\epsilon}.
\label{QM8}
\end{equation}
How could we estimate the probability distribution from actual data? Is it somehow related to the statistics of cold and hot spots?
Leaving aside these important questions it is clear that the correct limit where Eq. (\ref{QM8}) should be 
analyzed is given by $n \gg 1$ and $\overline{n}\gg 1$. More specifically, the realizations of the distribution (\ref{QM8}) 
with different $\overline{n}$. In the limit $\overline{n} \gg \epsilon$  the negative binomial distribution becomes a Gamma distribution
with a characteristic scaling law:
\begin{equation}
\overline{n} P_{n}(\overline{n}, \epsilon) \simeq \psi_{\epsilon}(w), \qquad 
\psi_{\epsilon}(w)= \frac{\epsilon^{\epsilon}}{\Gamma(\epsilon)} w^{\epsilon-1} e^{-\epsilon w},\qquad w= n/\overline{n}.
\label{av14}
\end{equation} 
discussed in quantum optics and also relevant to in high-energy physics \cite{KNO}. The scaling limit is defined by the asymptotic behaviour of $P_{n}(\overline{n})$ as $n \to \infty$,
 $\overline{n} \to \infty$ while $n/\overline{n}$ is kept fixed. For fixed $\epsilon$ the distribution 
 of Eq. (\ref{QM8}) scales as $\psi_{\epsilon}(w)$. 
\newpage
\renewcommand{\theequation}{7.\arabic{equation}}
\setcounter{equation}{0}
\section{Concluding remarks}
\label{sec7}
There are some who think that the quantum fluctuations are the only plausible initial data for the evolution of the metric perturbations. The complementary approach pursued here has been to scrutinize the coherence properties of large-scale curvature inhomogeneities when the relevant wavelengths are larger than the Hubble radius. This step is both plausible and mandatory 
if we ought to distinguish the quantum nature of large-scale curvature inhomogeneities 
from more mundane statistical correlations.

Since the pioneering attempts of Hanbury Brown and Twiss it has been realized that the analysis of first order interference 
between the amplitudes cannot be used to distinguish the nature of different quantum states of the radiation field. Young interferometry is not able, by itself, to provide information on the statistical properties of the quantum correlations since various states with diverse physical properties (such as laser light and chaotic light) may lead to comparable degrees of first-order coherence. The Hanbury Brown-Twiss interferometry demand  the analysis of the second-order correlations of corresponding intensities. The degree of second-order coherence, however,  does not describe unambiguously the correlation properties of large-scale curvature perturbations. The applications of the Glauber theory allow for a systematic scrutiny of the higher-order correlations.  More specifically the analysis of the degrees of third- and fourth-order coherence is necessary to assess the statistical properties of curvature inhomogeneities and their plausible quantum origin. Similar conclusions have been recently drawn in a quantum optical context where higher-order autocorrelations are mandatory for characterizing the multiphoton nature of nonclassical light. 

Since we showed that the statistical properties of the quantum states can be disambiguated 
by examining the higher degrees of coherence, it would be interesting to tailor specific observational strategies along this direction.  
The new generations of CMB detectors and the Hanbury Brown-Twiss interferometry in the THz region 
could be a reasonable hope in this respect. While this possibility 
seems rather interesting, there are some who might object that an improved statistical accuracy in the determination of the parameters of the concordance paradigm should be the primary target of future experimental endeavours. A full account of this potentially interesting debate is beyond the scopes of the present investigation.
\section*{Acknowledgements}
It is a pleasure to thank T. Basaglia and S. Rohr of the CERN scientific information service for their kind assistance.  

\newpage

\begin{appendix}
\renewcommand{\theequation}{A.\arabic{equation}}
\setcounter{equation}{0}
\section{Hamiltonians for gravitons and phonons}
\label{APPA}
The  essentials of the quantum description of the scalar and tensor modes of the geometry will be swiftly summarized in terms of 
the following Hamiltonian:
\begin{equation}
\hat{H}(\tau) = \frac{1}{2} \int d^{3} x \biggl[ \hat{\pi}^2 - 2\, i\,  \lambda ( \hat{\pi} \hat{q} + 
\hat{q} \hat{\pi}) + \partial_{k} \hat{q} \partial^{k} \hat{q}\biggr],
\label{AA1}
\end{equation}
where $\hat{q}$ and $\hat{\pi}$ are the two canonically conjugate field operators and $\lambda$ is the pump field. The specific expressions of $\hat{q}$, $\hat{\pi}$ and $\lambda$ differ in the case of the scalar and tensor modes of the geometry; the form of Eq. (\ref{AA1}) remains however the same in both cases. In the tensor case  $\hat{q}(\vec{x},\tau)$ is given by: 
\begin{equation}
\hat{q}_{t}(\vec{x},\tau) = a(\tau) \, \hat{h}(\vec{x},\tau), \qquad \hat{\pi}_{t} = \hat{q}_{t}^{\prime} - {\mathcal H} \hat{q}_{t},\qquad \lambda\to \lambda_{t} = \frac{i}{2} {\mathcal H} = \frac{a^{\prime}}{a}, 
\label{AA1a}
\end{equation}
where $\hat{h}(\vec{x},\tau)$ is the operator corresponding to a single tensor polarization, $a(\tau)$ is the scale factor and the prime denotes a derivation with respect to $\tau$.  In the scalar case we have instead: 
\begin{equation}
\hat{q}_{s}(\vec{x},\tau) = z(\tau) \, \hat{{\mathcal R}}(\vec{x},\tau), \qquad \hat{\pi}_{s} =  \hat{q}_{s}^{\prime} - \frac{z^{\prime}}{z} \hat{q}_{s},\qquad \lambda\to \lambda_{s} = \frac{i}{2}  \frac{z^{\prime}}{z},
\label{AA1b}
\end{equation}
where ${\mathcal R}(\vec{x},\tau)$ denotes the (gauge-invariant) curvature perturbation\footnote{In this paper 
the tensor fluctuations of the geometry are defined as $\delta_{(\mathrm{t})} g_{ij} = - a^2 h_{ij}$ (subjected to the conditions $\partial_{i}h^{i}_{j} = h_{i}^{i} =0$) \cite{fordp};
the notation $\delta_{(\mathrm{t})}$ describes the tensor fluctuation of the metric. Even if the curvature perturbations are invariant under infinitesimal coordinate transformation (as implied by the Bardeen formalism \cite{bard1}) their explicit expression changes from one coordinate system to the other. 
In the conformally Newtonian gauge the scalar modes of the geometry are given by $\delta_{(\mathrm{s})} g_{00} = 2 a^2 \phi$ and $\delta_{(\mathrm{s})} g_{ij} = 2 a^2 \psi \, \delta_{ij}$,
where $\delta_{(\mathrm{s})}$ denotes, respectively, the scalar fluctuation of the corresponding entry of the metric. In this frame the curvature perturbations are defined as ${\mathcal R} = - \psi - ({\mathcal H}/\varphi^{\prime}) \chi$ 
where $\psi$ is the longitudinal fluctuation of the metric and $\chi$ is the fluctuation of the inflaton $\varphi$. }  on comoving orthogonal hypersurfaces; note that  $z = a \varphi^{\prime}/{\mathcal H}$ where $\varphi$ is the background scalar field (not necessarily identified with the inflaton).
In what follow we shall simply posit $2 \lambda= i \,{\mathcal F}$ where ${\mathcal F}$ can be alternatively identified either with ${\mathcal H}$ or with $z^{\prime}/z$. Note, in this respect, that $\lambda^{*} = - \lambda$ as required by the Hermiticity of the Hamiltonian given in Eq. (\ref{AA1}). 

The action for the normal mode of the tensor fluctuations of the geometry has been discussed for the first time by Ford and Parker \cite{fordp}. The gauge-invariant action for the scalar fluctuations (i.e. scalar phonons) appeared in a paper by Lukash \cite{luk} in the context 
of fluid models. Later on different authors applied it to scalar field matter with particular attention to the 
quantization of the fluctuations \cite{KS,chibisov}.  The Lukash variable coincides, in the scalar field case, with curvature perturbation 
on comoving orthogonal hypersurfaces. A thorough discussion of the Hamiltonians for the scalar and tensor modes and of their quantization 
can be found in \cite{mg1}. Note, in particular, that since the Hamiltonian (\ref{AA1}) is time dependent, it is always possible to perform time-dependent canonical transformations, leading to a different Hamiltonian; by definition the evolution equations of the field operators will remain the same. This potential 
ambiguity will play no role in our considerations but can have some impact on the specific form of the initial state. The Fourier representation of the field operators can be written as: 
\begin{eqnarray}
\hat{q}(\vec{x},\tau) &=& \frac{1}{\sqrt{V}} \sum_{\vec{p}} \hat{q}_{\vec{p}}(\tau)\, e^{- i \vec{p}\cdot \vec{x}},\qquad 
\hat{\pi}(\vec{x},\tau) = \frac{1}{\sqrt{V}} \sum_{\vec{p}} \hat{\pi}_{\vec{p}}(\tau)\, e^{- i \vec{p}\cdot \vec{x}},
\label{AA2}\\
\hat{q}_{\vec{p}} &=& \frac{1}{\sqrt{2 p}} ( \hat{a}_{\vec{p}} + \hat{a}_{-\vec{p}}^{\dagger}),\qquad \hat{\pi}_{\vec{p}} = - i \sqrt{\frac{p}{2}} ( \hat{a}_{\vec{p}} - \hat{a}_{-\vec{p}}^{\dagger}),
\label{AA3}
\end{eqnarray}
where $V$ represents a fiducial (normalization) volume. In the discussion it is practical 
to switch from discrete to continuous modes where the creation 
and annihilation operators obey $[\hat{a}_{\vec{k}}, \hat{a}_{\vec{p}}^{\dagger}] = 
\delta^{(3)}(\vec{k} - \vec{p})$ and the sums are replaced by integrals according to 
$\sum_{\vec{k}} \to V \int d^{3} k/(2\pi)^3$. This simple observation should be borne in mind when discussing 
the explicit results. In terms of Eqs. (\ref{AA2}) and (\ref{AA3}) the Hamiltonian of Eq. (\ref{AA1}) becomes:
\begin{equation}
\hat{H}(\tau) = \int d^{3}k \biggl[ \frac{k}{2} (\hat{a}_{\vec{k}}^{\dagger} \, \hat{a}_{\vec{k}} + \hat{a}_{-\vec{k}} \, \hat{a}_{-\vec{k}}^{\dagger})
+ \lambda\, \hat{a}_{\vec{k}}^{\dagger} \hat{a}_{-\vec{k}}^{\dagger}  + \lambda^{\ast} \,  \hat{a}_{-\vec{k}}\, \hat{a}_{\vec{k}}\biggr].
\label{AA4}
\end{equation} 
The system of Eq. (\ref{AA4}) has been firstly discussed by Mollow and Glauber \cite{sq1} in the quantum theory of parametric amplification\footnote{ Equation (\ref{AA4}) describes an interacting bose gas at zero temperature. In this case the free 
Hamiltonian corresponds to the kinetic energy while the interaction terms account for the two-body collisions with small momentum transfer  \cite{fetter,solomon}.}. 

\renewcommand{\theequation}{B.\arabic{equation}}
\setcounter{equation}{0}
\section{Diagonalization of the Hamiltonian}
\label{APPB}
The initial conditions of curvature perturbations are fixed by imposing that at the onset of the dynamical evolution 
(i.e. at some initial time $\tau_{i}$) the Hamiltonians belonging to the class of Eq. (\ref{AA4}) are separately minimized\footnote{The initial conditions might not be quantum mechanical \cite{mg1}, as discussed in the bulk of the paper.}. To diagonalize
the Hamiltonian of Eq. (\ref{AA4}) we first perform the following canonical transformation\footnote{Note that $u_{p}(\tau)$ and $v_{p}(\tau)$ must obviously satisfy $|u_{p}(\tau)|^2 - |v_{p}(\tau)|^2 =1$. The two complex functions $u_{p}(\tau)$ and $v_{p}(\tau)$ can the be parametrized in terms of three real functions.} :
\begin{equation}
\hat{a}_{\vec{p}}(\tau,\,\tau_{i}) = u_{p}(\tau) \hat{b}_{\vec{p}}(\tau_{i}) - v_{p}(\tau) \hat{b}_{- \vec{p}}^{\dagger}(\tau_{i}),\qquad 
\hat{a}_{-\vec{p}}^{\dagger}(\tau,\,\tau_{i}) = u_{p}^{\ast}(\tau) \hat{b}_{-\vec{p}}^{\dagger}(\tau_{i}) - v_{p}^{\ast}(\tau) \hat{b}_{\vec{p}}(\tau_{i}).
\label{AA5}
\end{equation}
 Inserting Eq. (\ref{AA5}) into Eq. (\ref{AA4}) the following equation is obtained, after some algebra,
\begin{eqnarray}
\hat{H}(\tau) &=& \int d^{3} k \biggl\{ \biggl[ \frac{k}{2} \biggl( |u_{k}(\tau)|^2 +  |v_{k}(\tau)|^2 \biggr)- \lambda(\tau) u_{k}^{\ast}(\tau) v_{k}^{\ast}(\tau)  
\nonumber\\
&-&  \lambda^{\ast}(\tau) u_{k}(\tau) v_{k}(\tau)  \biggr] ( \hat{b}_{\vec{p}}^{\dagger}\, \hat{b}_{\vec{p}}+ \hat{b}_{-\vec{p}}\, \hat{b}_{-\vec{p}}^{\dagger})
\nonumber\\
&+& \biggl[ \lambda u_{k}^{\ast\,2}(\tau) + \lambda^{\ast} v_{k}^{2} - \frac{k}{2} (u_{k}^{\ast}(\tau) v_{k}(\tau) + u_{k}(\tau) v_{k}^{\ast}(\tau))\biggr]
 \hat{b}_{\vec{p}}^{\dagger}\, \hat{b}_{-\vec{p}}^{\dagger}
\nonumber\\
&+& \biggl[ \lambda^{\ast} u_{k}^{2}(\tau) + \lambda v_{k}^{\ast\,2} - \frac{k}{2} (u_{k}(\tau) v_{k}^{\ast}(\tau) + u_{k}^{\ast}(\tau) v_{k}(\tau))\biggr]
 \hat{b}_{-\vec{p}}\, \hat{b}_{\vec{p}}\biggr\}.
\label{AA6}
\end{eqnarray}
To diagonalize the Hamiltonian we must require that the third and fourth lines at the right hand side of Eq. (\ref{AA6}) 
disappear for $\tau = \tau_{i}$. This happens when the condition $\lambda \, u_{k}^{\ast\, 2} + \lambda^{\ast} v_{k}^{2} = k u^{\ast}_{k} v_{k}$ 
is verified\footnote{From now on it is understood that the various functions must be evaluated for $\tau= \tau_{i}$ so that, for the moment, the arguments 
of the various functions shall be omitted. }
Since $u_{k}$ and $v_{k}$ are two complex variables subjected to one condition valid for any $\tau$ they can be parametrized 
in terms of three real quantities as:
\begin{equation}
u_{k} = e^{i \theta_{k}} \cosh{r_{k}}, \qquad v_{k} = e^{i \beta_{k}} \sinh{r_{k}},\qquad \lambda = |\lambda| e^{i \delta},
\label{AA8}
\end{equation}
where $\delta$ is $k$-independent. 
Using Eq. (\ref{AA8}) the relation $\lambda \, u_{k}^{\ast\, 2} + \lambda^{\ast} v_{k}^{2} = k u^{\ast}_{k} v_{k}$ becomes:
\begin{equation}
|\lambda| [ \cosh^2{r_{k}} + e^{ 2 i (\theta_{k} + \beta_{k} - \delta)} \sinh^2{r_{k}}] = k e^{ i (\theta_{k} + \beta_{k} - \delta)} \cosh{r_{k}} \, \sinh{r_{k}}.
\label{AA9}
\end{equation}
Separating the conditions for the real and the imaginary parts we have 
\begin{eqnarray}
&& |\lambda| [\cosh^2{r_{k}} + \cos{2 (\theta_{k} +\beta_{k} - \delta)} \sinh^2{r_{k}}]  = k  \cos{(\theta_{k} +\beta_{k} - \delta)} 
\cosh{r_{k}} \sinh{r_{k}},
\nonumber\\
&& |\lambda| \sinh^2{r_{k}} \sin{2(\theta_{k} +\beta_{k} - \delta)} = k  \sin{(\theta_{k} +\beta_{k} - \delta)} 
\cosh{r_{k}} \sinh{r_{k}}. 
\label{AA10}
\end{eqnarray}
Both conditions in Eq. (\ref{AA10}) can be satisfied when the conditions $2 |\lambda| \cosh{2 r_{r}} =  k\,\sinh{2 r_{k}}$ and $\theta_{k} + \beta_{k} = \delta$ are simultaneously satisfied.
Since $2 \lambda = i {\mathcal H}$ (or $2\lambda = i z^{\prime}/z$) we can easily see that, in the realistic situation,
 $\delta = \pi/2$ (i.e. $\lambda= - \lambda^{*}$) and therefore 
 \begin{equation}
 \tanh{2 r_{k}} = \frac{ 2 |\lambda(\tau_{i})|}{k}, \qquad \theta_{k} + \beta_{k} =i.
 \label{AA12}
 \end{equation}
 But since the first relation at the right hand side lies between $-1$ and $1$ this equation can be solved for all $k$ 
 provided $|\lambda(\tau_{i})/k| \leq 1/2$. But since $|\lambda(\tau_{i})| = 1/(2 \tau_{i})$ this simply means that the 
 relevant modes must be all inside the Hubble radius at $\tau_{i}$, i.e. $k \tau_{i} \gg 1$.
 The diagonalized Hamiltonian at $\tau_{i}$ will then be\footnote{While in flat space the zero-point energy might be renormalized, 
  in the present case the energy density of the initial state must not exceed the energy density of the background. This is what happens for quantum mechanical initial data. See however Refs. \cite{mg1} and 
\cite{max1} for further discussions on this theme in more general situations.}:
 \begin{equation}
 \hat{H}(\tau_{i}) = \int d^{3}k \, \frac{k}{2 \cosh{2 r_{k}}} ( \hat{b}_{\vec{p}}^{\dagger}\, \hat{b}_{\vec{p}}+ \hat{b}_{-\vec{p}}\, \hat{b}_{-\vec{p}}^{\dagger}).
 \label{AA13}
 \end{equation}
 
\renewcommand{\theequation}{C.\arabic{equation}}
\setcounter{equation}{0}
\section{Quantum theory of parametric amplification}
\label{APPC} 
From the Hamiltonian (\ref{AA4}) the evolution equations in the Heisenberg description can be easily derived and they are given by: 
\begin{equation}
\frac{d \hat{a}_{\vec{p}}}{d\tau} = - i\, p \, \hat{a}_{\vec{p}} - 2 i \, \lambda \hat{a}_{- \vec{p}}^{\dagger},\qquad \frac{d \hat{a}_{\vec{p}}^{\dagger}}{d\tau} =  i\, p \, \hat{a}_{\vec{p}} + 2 i \, \lambda^{\ast} \hat{a}_{\vec{p}}.
\label{AA14}
\end{equation}
From Eq. (\ref{AA14}) the equations obeyed by $u_{p}$ and $v_{p}$ can be written as:
\begin{equation}
\frac{d u_{p}}{d\tau} = - i p \,u_{p} - {\mathcal F} v_{p}^{\ast}, \qquad \frac{d v_{p}}{d\tau} = - i p\, v_{p} - {\mathcal F} u_{p}^{\ast},
\label{AA15}
\end{equation}
where, as explained in appendix \ref{APPA}, ${\mathcal F}$ denotes either ${\mathcal H}$ or $z^{\prime}/z$. If the fluctuations evolve in a quasi-de Sitter space-time the functions  $u_{k}(\tau)$ and $v_{k}(\tau)$ can be expressed as\footnote{The mode functions $f_{p}$ and $g_{p}$ are related to $u_{p}$ and $v_{p}$ as $u_{p} - v_{p}^{\ast} = \sqrt{2 p} \, f_{p}$ and  as $u_{p} + v_{p}^{\ast} = i \sqrt{2/p} g_{p}$. By construction it turns out that $g_{p} = f_{p}^{\prime} - {\mathcal F} f_{p}$ (see also \cite{max1} for details).}
\begin{eqnarray}
u_{k}(\tau) &=& - \frac{i}{2} {\mathcal N}\, \sqrt{- k\tau} \biggl[ H_{\mu-1}^{(1)}(- k\tau) + i \, H_{\mu}^{(1)}(- k\tau) + \frac{3 - 2 \mu}{2(- k\tau)} H_{\mu}^{(1)}(- k\tau)\biggr],
\nonumber\\
v_{k}(\tau) &=&  - \frac{i}{2} {\mathcal N}^{*} \, \sqrt{- k\tau} \biggl[ H_{\mu-1}^{(2)}(- k\tau) - i \, H_{\mu}^{(2)}(- k\tau) + \frac{3 - 2 \mu}{2(- k\tau)} H_{\mu}^{(2)}(- k\tau)\biggr],
\label{AA18}
\end{eqnarray}
where ${\mathcal N} = \exp{[i \pi( \mu + 1/2)/2]}$.
The explicit values of $\mu$ differ slightly in the tensor and in the scalar case, at least in the context of conventional (i.e. single-field)
inflationary scenarios:
\begin{equation}
\mu_{tensor} = \frac{3 - 2 \epsilon}{2( 1 - \epsilon)}, \qquad \mu_{scalar} = \frac{3 + \epsilon + 2\eta}{2 (1 -\epsilon)},
\label{AA18a}
\end{equation}
where $\mu_{tensor}$ and $\mu_{scalar}$ denote, respectively, the values of $\mu$ in the tensor and in the scalar cases;
Eq. (\ref{AA18}) evaluated in the limit $\mu \to 3/2$ implies: 
implies 
\begin{equation}
u_{k}(\tau) = e^{- i k\tau} \biggl( 1 - \frac{i}{2 k\tau}\biggr), \qquad v_{k}(\tau) = - \frac{i}{2 k \tau} e^{i k\tau}.
\label{AA19}
\end{equation}

The evolution of $u_{k}$ and $v_{k}$ can be rephrased in terms of the so-called squeezing parameters \cite{sq2,sq3} (see also \cite{max1,gr1,mg2}). 
More specifically Eq. (\ref{AA5}) can be rewritten as\footnote{In the following formulae $\varphi_{p}$, $r_{p}$ and $\gamma_{p}$ are all functions of the 
conformal time coordinate $\tau$ even if the arguments of the functions will be dropped for the sake of conciseness.}
\begin{equation} 
\hat{a}_{\vec{p}} = e^{- i \varphi_{p}}\biggl[ \cosh{r_{p}}\, \hat{b}_{\vec{p}} - e^{i \gamma_{p}} \sinh{r_{p}} \hat{b}^{\dagger}_{-\vec{p}}\biggr],\qquad
\hat{a}_{-\vec{p}}^{\dagger} =  e^{i \varphi_{p}}\biggl[ \cosh{r_{p}}\, \hat{b}_{-\vec{p}}^{\dagger} - e^{-i \gamma_{p}} \sinh{r_{p}}\hat{b}_{\vec{p}}\biggr],
\label{AA20}
\end{equation}
implying that $u_{k}=  e^{- i \varphi_{k}}\cosh{r_{k}}$ and $v_{k}= e^{- i (\varphi_{k} -\gamma_{k})} \sinh{r_{k}}$. This parametrization makes 
explicit the dependence of the two complex functions $u_{k}$ and $v_{k}$ in terms of three real functions of the conformal time coordinate $\tau$.  
Using Eq. (\ref{AA20}) into  Eqs. (\ref{AA14})--(\ref{AA15}), the evolution of the squeezing amplitude $r_{k}$ and of the two phases becomes:
\begin{equation}
r_{k}^{\prime} = - {\mathcal F} \cos{\alpha_{k}}, \qquad 
\varphi_{k}^{\prime} = k + {\mathcal F} \sin{\alpha_{k}}\, \tanh{r_{k}},\qquad 
\gamma_{k}^{\prime} = \varphi_{k}^{\prime} - k - {\mathcal F} \frac{\sin{\alpha_{k}}}{\tanh{r_{k}}},
\label{AA24}
\end{equation}
where $\alpha_{k} = 2 \varphi_{k} - \gamma_{k}$;  combining the previous equations it is also simple to obtain
$\alpha_{k}^{\prime} = 2 k + 2 {\mathcal F} \sin{\alpha_{k}}/\tanh{2 r_{k}}$ (see also Refs. \cite{max1,mg1}). 

Equation (\ref{AA20}) can be swiftly obtained by considering a single $\vec{p}$-mode and by noticing that 
the operators ${\mathcal K}_{\pm}$ and ${\mathcal K}_{0}$ obey the commutation relations of the $SU(1,1)$ Lie algebra:
\begin{equation}
{\mathcal K}_{+}= \hat{b}_{1}^{\dagger} \,\hat{b}_{2}^{\dagger},\qquad {\mathcal K}_{-} = \hat{b}_{1}\, \hat{b}_{2},
\qquad {\mathcal K}_{0}= \frac{1}{2}\biggl[ \hat{b}_{1}^{\dagger}\, \hat{b}_{1} + \hat{b}_{2} \,\hat{b}_{2}^{\dagger}\biggr].
\label{AA21}
\end{equation}
Using the the standard Campbell-Baker-Hausdorff theorem \cite{mandel}, Eq. (\ref{AA21}) implies 
\begin{equation}
\hat{a} = \Sigma^{\dagger}(\zeta) \, \Xi^{\dagger}(\varphi) b_{1} \Xi(\varphi) \Sigma(\zeta) = 
e^{- i \varphi}\biggl[ \cosh{r}\, \hat{b}_{1} - e^{i \gamma} \sinh{r} \hat{b}^{\dagger}_{2}\biggr],
\label{AA22}
\end{equation}
where $\Xi(\varphi)$ and $\Sigma(\zeta)$ (with $\zeta= r e^{i\gamma}$) are, respectively, the rotation operator and the two-mode 
squeezing operators \cite{sq2,sq3} defined in terms of the generators of the $SU(1,1)$ Lie algebra:
\begin{equation}
\Xi(\varphi) = \exp{[ - i \varphi (\hat{b}_{1}^{\dagger} \hat{b}_{1} + \hat{b}_{2} \hat{b}_{2}^{\dagger})]}, \qquad \Sigma(\zeta)= 
\exp{[ \zeta^{*} \hat{b}_{1} \hat{b}_{2} - \zeta \hat{b}_{2}^{\dagger} \hat{b}_{1}^{\dagger}]}.
\label{AA23}
\end{equation}
These two operators play an important role in the evolution of the states in the Schr\"odinger representation; their use has been 
pioneered by Grishchuk and Sidorov \cite{gr1} (see also \cite{mg2}). 

We conclude by proving that the normal-ordered expectation values dominate the degrees of coherence provided the average multiplicity of the produced phonons and gravitons is large; this has been dubbed in the bulk of the paper as the $1/\overline{n}$ expansion. Consider, for this purpose, the following normalized expectation values:
\begin{equation}
\overline{g}^{(N)} = \frac{\langle \, \hat{b}^{\dagger}\, \hat{b}^{\dagger} \, \hat{b}\, \hat{b} \rangle}{\langle \, \hat{b}^{\dagger} \, \hat{b}\, \rangle^2}, \qquad 
\overline{g}^{(A)} = \frac{\langle \, \hat{b}\, \hat{b}\, \hat{b}^{\dagger} \, \hat{b}^{\dagger}\, \rangle}{\langle \, \hat{b}^{\dagger} \, \hat{b}\, \rangle^2},
\qquad \overline{g}^{(S)} = \frac{\langle \, \hat{b}\, \hat{b}^{\dagger} \, \hat{b}^{\dagger} \, \hat{b}\, \rangle}{\langle \, \hat{b}^{\dagger} \, \hat{b}\, \rangle^2},
\label{CORRz}
\end{equation}
where the superscripts  denote, respectively, the normal, antinormal and symmetric ordering. Using the commutation relations the first two relations of Eq. (\ref{CORRz}) can be written in terms of the normal-ordered correlator. For the antinormal correlator we have 
$\overline{g}^{(A)} = \overline{g}^{(N)} +3/\overline{n}+ 2/\overline{n}^2$ while for the symmetric correlator we get 
$\overline{g}^{(S)} = \overline{g}^{(N)} + 2/\overline{n}$ where, as usual, $\overline{n} = \langle \hat{N}\rangle$ is the average multiplicity. Since $\overline{n} \gg 1$ all the terms ${\mathcal O}(1/\overline{n})$ can be neglected. In quantum optics 
it is natural to impose the normal ordering in the higher-order correlators since the detection of light quanta (i.e. in the optical range of frequencies) occurs by measuring a photo-current, i.e. a current induced by the absorption of a photon. Notice finally that using the commutation relations we can easily obtain $\overline{g}^{(N)} = 1 + (D^2 - \overline{n})/\overline{n}^2$ where $D^2= \langle \, \hat{N}^2\, \rangle - \langle\, \hat{N}\,\rangle^2$ denotes in the bulk of the paper the variance of the underlying multiplicity distribution. 

\end{appendix}

\end{document}